\documentclass[12pt,aps]{revtex4}

\usepackage{bbm}
\usepackage{amssymb}
\usepackage{amsmath}

\newcommand{\nc}{\newcommand}
\nc{\be}{\begin{equation}}
\nc{\ee}{\end{equation}}
\nc{\bea}{\begin{eqnarray}}
\nc{\eea}{\end{eqnarray}}
\nc{\nn}{\nonumber}
     
\nc{\lp}{\left(}
\nc{\rp}{\right)}

\catcode`\*=11  
\def\slashsym#1#2{\mathpalette{\sl*sh{#1}}{#2}}
\def\sl*sh#1#2#3{\ooalign{\setbox0=\hbox{$#2\not$}
                          $\hfil#2\mkern-24mu\mkern#1mu
                           \raise.15\ht0\box0\hfil$\cr
                          $#2#3$}}
\catcode`\*=12
\def\Eins{\mathbbm{l}}

\def\pslash{{\slashsym8p}}

\def\Aslash{{\slashsym9A}}

\def\Bslash{{\slashsym6B}}

\def\dslash{{\slashsym5\partial}}

\def\Ac{{\mathcal{A}}}

\def\Fc{{\mathcal{F}}}
\def\Hc{{\mathcal{H}}}
\def\Kc{{\mathcal{K}}}
\def\Dc{{\mathcal{D}}}
\def\Pc{{\mathcal{P}}}
\def\Lc{{\mathcal{L}}}
\def\Oc{{\mathcal{O}}}

\def\Nc{{\mathcal{N}}}
\def\Ocd{{\mathcal{O}^{\dagger}}}
\def\Xc{{\mathcal{\chi}}}
\def\Det{{\mathrm{Det\,}}}
\def\Tr{{\mathrm{Tr\,}}}
\def\Ub{{\mathbf{U}}}

\nc{\markb}{$\clubsuit \Rightarrow $}
\nc{\marke}{$ \Leftarrow \clubsuit $}
\nc{\markx}{$ \clubsuit !! \clubsuit $}
\nc{\markeq}{ \clubsuit !! \clubsuit }

\nc{\eq}{{Eq.}}
\nc{\eqs}{{Eqs.}}
\nc{\tr}{\textrm{tr}}

\begin{document}
\rightline{HD-THEP-07-19}


\title{Effective Action in a General Chiral Model: Next to Leading 
Order Derivative Expansion in the Worldline Method}

\author{Andres Hernandez$^{(1)}$}
\email[]{A.Hernandez@thphys.uni-heidelberg.de}

\author{Thomas Konstandin$^{(2)}$}
\email[]{Konstand@kth.se}

\author{Michael G. Schmidt$^{(1)}$}
\email[]{M.G.Schmidt@thphys.uni-heidelberg.de}

\affiliation{$(1)\,$
Institut f\"ur Theoretische Physik, Heidelberg
University, Philosophenweg 16, D-69120 Heidelberg, Germany
}%

\affiliation{$(2)\,$Department of Theoretical Physics, Royal Institute of Technology (KTH), 
        AlbaNova University Center, Roslagstullsbacken 21, 106 91
        Stockholm, Sweden}

\date{\today}

\begin{abstract}
We present a formalism to determine the imaginary part of a general
chiral model in the derivative expansion. Our formalism is based on
the worldline path integral for the covariant current that can be
given in an explicit chiral and gauge covariant form. The effective
action is then obtained by integrating the covariant current, taking
account of the anomaly.
\end{abstract}

\maketitle

%
%


\section{Introduction}

When discussing the influence of fermions on the dynamics of some
theory, e.g. of the Standard Particle Model (SM) in a cosmology
setting, it is mandatory to integrate out the fermions. This procedure
in one-loop order already becomes quite involved if arbitrary chiral
couplings of space-time dependent outer bosonic fields are
considered. The extensive work of Salcedo \cite{Salcedo1,Salcedo2}
resulted in an effective action to leading order in covariant
derivatives of such fields. Basis is a refined calculation in momentum
space, handling of anomalies using the Wess-Zumino-Witten model (WZW),
and last but not least, a very practical shorthand notation.

Such effective actions are quite important in evaluating some models.
For example, in ref.~\cite{Smit} Smit discussed a form of 'cold'
electroweak baryogenesis at the end of electroweak scale
inflation~\cite{Tranberg} which could very well work if the rephasing
invariant
\be
J=s_1^2s_2s_3c_1c_2c_3 \sin(\delta)=(3.0\pm0.3)\times10^{-5}
\ee
of the Jarlskog determinant is not accompanied by further suppressions
through mass ratios. It was proposed \cite{Smit} that derivative terms
in the effective action that are analytic in the time-dependent masses
considered non-perturbatively could be very important in
non-equilibrium. Such effects were also observed in
ref.~\cite{Konstandin2}. In the work~\cite{Smit} the fourth order
derivative result of ref.~\cite{Salcedo2} turned out not to contain
CP violation, but the claim was made that higher orders of the
imaginary part of the effective action will do.

Worldline methods in first quantized quantum field theory are ideally
adapted for calculating effective actions: One considers the
propagation of a particle in some space-time dependent background
\cite{Schwinger}, but in x-space path integral formulation
\cite{Feynman}. This method~\cite{Strassler, Schmidt4}, also related to the infinite
tension limit of String theory \cite{Bern1,Bern2}, was used heavily
for the discussion of various effective actions in one-loop
\cite{Schmidt4,Schmidt5,Schmidt6,Schmidt1} and two-loop
\cite{Schmidt2loop2,Schmidt2loop,Schmidt3,TwoQCD} order. For example, the high
order in the inverse mass calculation of ref.~\cite{Schmidt3} could
hardly be done with other methods.

The present paper provides a formalism to determine higher order
contributions to the imaginary part of the effective action using the
worldline formalism. We are concerned with the effective action of a
multiplet of N Dirac fermions coupled to an arbitrary matrix-valued
set of fields, including a scalar $\Phi$, a pseudoscalar $\Pi$, a
vector \textit{A}, a pseudovector \textit{B}, and an antisymmetric
tensor \textit{$K_{\mu \nu}$}. One peculiar feature of the imaginary
part of the effective action is that it cannot be written in a
manifest chiral covariant way, due to the presence of the chiral
anomaly. One possibility to arrive at a closed expression for the
effective action is to abandon manifest chiral covariance as it was
done in ref.~\cite{Gagne2}. The resulting expression is rather
complicated and not well suited for higher order
calculations. Alternatively, it was proposed in ref.~\cite{Salcedo2}
to determine the covariant current for which a manifestly chiral
covariant expression exists and to take account of the anomaly when
integrating the current to yield the effective action. Following this
idea, we present a worldline path integral formulation of the
covariant current.

Before we do so, we review the worldline formalism by discussing the
derivation of the real part of the effective action. A single Dirac
fermion in the presence of both a scalar and pseudoscalar fields in
the context of the worldline formalism was first treated in
ref.~\cite{Axial1}, the inclusion of a pseudovector in
ref.~\cite{Axial2}. In our discussion of the real part of the
effective action we will follow the elegant subsequent work of
refs.~\cite{Gagne1,Gagne2}.

Section \ref{sec_act} contains the derivation of the real part of the
effective action, the derivation of the covariant current and the
matching procedure to obtain the imaginary part of the effective
action. In section \ref{sec_low}, we briefly reproduce the results in
lowest order from ref.~\cite{Salcedo2}. As a novel result we present
the imaginary part of the effective action in two dimensions in next
to leading order in section~\ref{sec_next}.

\section{Effective Action\label{sec_act}}

We are concerned with the effective action 
\begin{equation}
\label{W}
i\, W [ \Phi,\Pi,A,B,K ]=\log \Det i\, [ \, i\, \dslash -\Phi + 
\, i\,  \, \gamma^5 \Pi + \Aslash + \gamma^5 \Bslash + 
\, i \, \gamma^{\mu} \gamma^{\nu} K_{\mu \nu} ],
\end{equation}
and its continuation to Euclidean space. The $\gamma$ matrices remain
unaffected by the continuation, but it is useful to introduce the
following notation, $(\gamma_E)_j
\equiv i \gamma_j$, $(\gamma_E)_4 \equiv
\gamma_0$, and $(\gamma_E)_5 \equiv \gamma_5$. 
After Wick-rotation, $t\rightarrow -it$, one obtains with this new
notation
\begin{equation}
\label{wickrot}
\dslash \rightarrow i \dslash_E, \, \Aslash \rightarrow 
i \Aslash_E, \, \Bslash \rightarrow i \Bslash_E, 
\, \gamma^{\mu} \gamma^{\nu} K_{\mu \nu} \rightarrow 
- (\gamma_E)_{\mu} (\gamma_E)_{\nu} K_{E \mu \nu}.
\end{equation} 
From now on, the $E$ subscript will be suppressed.

The effective action of \eq~(\ref{W}) now reads
\begin{equation}
\label{WE}
-W [ \Phi,\Pi,A,B,K ] = \log \Det [ \Oc ],
\end{equation} 
with the operator $\Oc$ in momentum space defined by 
\begin{equation}
\label{O}
\Oc \equiv \pslash - i \Phi(x) - \gamma_5 \Pi(x) - \Aslash(x) 
- \gamma_5 \Bslash(x) + \gamma_{\mu} \gamma_{\nu} K_{\mu \nu}.
\end{equation} 
As in ref.~\cite{Gagne1, Salcedo1}, the real and imaginary parts of
the effective action are analyzed separately
\begin{equation}
\label{Sep}
-W^{+}-i\,W^{-}=\log \left(\vert \Det[\Oc]\vert\right)+i\,\arg\left(\Det[\Oc]\right).
\end{equation} 
A perturbative expansion in weak fields \cite{Gagne1} shows that
graphs with an even number of $\gamma_5$ vertices are real, and graphs
with an odd number of $\gamma_5$ vertices are imaginary. This will
prove useful when the behavior of the effective action under
complex conjugation is explored later on.

\subsection{Real Part of the Effective Action}

Our intention is to obtain a worldline representation for the
effective action with manifest chiral and gauge invariance. This is
unproblematic for the real part, but it causes certain difficulties
for the imaginary part due to the chiral anomaly. In order to
familiarize the reader with the worldline method we review the
derivation for the real part in four dimensions as it was presented in
refs.~\cite{Gagne1,Gagne2}.

\subsubsection{Construction of a Positive Operator for the Real Part of the Effective Action}

In order to use the worldline formalism, one has to rewrite the
effective action in terms of a positive operator, thus obtaining
\begin{equation}
W^{+}=-\frac{1}{2}\log \Det [ \Oc^{\dagger} \Oc ].
\end{equation} 
The problem with this operator is that it contains terms linear in the
$\gamma$ matrices, what makes the transition to a path integral of
Grassman fields problematic. One way to avoid this problem is by
doubling the fermion system and exchanging the operator $\Oc$ for a
Hermitian operator $\Sigma$ yielding
\begin{equation}
\label{WR}
W^{+}=-\frac{1}{2}\log \Det [ \Oc^{\dagger} \Oc ] = -\frac{1}{4} \log \Det [ \Sigma^2 ],
\qquad \quad
\Sigma \equiv  
\begin{pmatrix} 
0 & \Oc \\
\Oc^{\dagger} & 0 \\
\end{pmatrix}.
\end{equation} 
Since $\Sigma$ is Hermitian, one can use the Schwinger integral
representation of the logarithm without any restrictions. One obtains
\begin{equation}
\label{WRint}
W^{+}=\frac{1}{4}\int_0^{\infty}\frac{dT}{T} \Tr \exp(-T \Sigma^2).
\end{equation} 
At this point, it is natural to introduce six $8\times8$ Hermitian
$\Gamma_A$ matrices. These matrices satisfy
$\lbrace\Gamma_A,\Gamma_B\rbrace=2\delta_{A B}$, with $A,B=1..6$ and
are defined as
\begin{equation}
\label{Gdef}
\Gamma_{\mu}= 
\begin{pmatrix} 
0 & \gamma_{\mu} \\
\gamma_{\mu} & 0 \\
\end{pmatrix},\quad 
\Gamma_5= 
\begin{pmatrix} 
0 & \gamma_5 \\
\gamma_5 & 0 \\
\end{pmatrix}, \quad
\Gamma_6=
\begin{pmatrix} 
0 & i\, \Eins_4 \\
-i\, \Eins_4 & 0 \\
\end{pmatrix}.
\end{equation}
For later use we also introduce the equivalent of $\gamma_5$,
\begin{equation}
\label{G7def}
\Gamma_7 = -i\,\prod_{A=1}^6 \Gamma_A= 
\begin{pmatrix}
\Eins_4 & 0 \\ 
0 & -\Eins_4 \\
\end{pmatrix},
\end{equation} 
and $\Gamma_7$ anticommutes with all other $\Gamma$ matrices.

Expressing $\Sigma$ in terms of these new matrices yields~\cite{Gagne2},
\begin{equation}
\label{Sigmanew}
\Sigma=\Gamma_{\mu}(p_{\mu}-A_{\mu})-\Gamma_6 \Phi-\Gamma_5 \Pi
-i\, \Gamma_{\mu} \Gamma_5 \Gamma_6 B_{\mu}
-i\, \Gamma_{\mu} \Gamma_{\nu} \Gamma_6 K_{\mu \nu}.
\end{equation} 
The aim is to turn \eq~(\ref{Sigmanew}) into an expression which is
manifestly chiral covariant. This can be achieved by changing to a
basis in which $i\, \Gamma_5 \Gamma_6$ is diagonal~\cite{Gagne1} using
the following transformation
\begin{equation}
\label{Change}
M^{-1}i\Gamma_5\Gamma_6 M= 
\begin{pmatrix} 
\Eins_4 & 0 \cr 0 & -\Eins_4 \\
\end{pmatrix}, \qquad M= 
\begin{pmatrix}
\Eins_2 & 0 & 0 & 0 \\
0 & 0 & 0 & \Eins_2 \\
0 & 0 & \Eins_2 & 0 \\
0 & \Eins_2 & 0 & 0 \\
\end{pmatrix}.
\end{equation}
In this basis, $\Sigma$ takes the form
\begin{equation}
\label{newSigma}
\tilde{\Sigma} = M^{-1} \Sigma M = 
\begin{pmatrix}
\gamma_{\mu}(p_{\mu}-A_{\mu}^L) & \gamma_5(-i\, H+\frac{1}{2} 
\gamma_{\mu} \gamma_{\nu} K_{\mu \nu}^s) \\
-\gamma_5(-i\, H^{\dagger}+\frac{1}{2} \gamma_{\mu} \gamma_{\nu} 
K_{\mu \nu}^{s \dagger}) & \gamma_{\mu}(p_{\mu}-A_{\mu}^R) \\
\end{pmatrix},
\end{equation} 
which is manifestly chiral covariant. Here $A^L=A+B$, $A^R=A-B$,
$H=\Phi-i\,\Pi$, $K^s=K-i\,\tilde{K}$ and $\tilde K_{\mu\nu}=\frac12
\epsilon_{\mu\nu\rho\sigma} K^{\rho\sigma}$ have been defined.

The square of $\tilde{\Sigma}$ constitutes a positive operator which
is suitable for the worldline formalism. However, even though this
expression contains only even combinations of $\gamma$ matrices, the
coherent state formalism cannot yet be used to transform this
expression into a fermionic path integral. In the coherent state
formalism, the $\gamma_5$ matrices have to be rewritten as a product
of the other $\gamma$ matrices, what would result again in odd
combinations. One possible solution of this problem is to enlarge the
Clifford space, replacing the $\gamma$ matrices by $\Gamma$ matrices
\be
\gamma_A \to \Gamma_A = \gamma_A \otimes
\begin{pmatrix}
0 & 1 \\ 
1 & 0 \\
\end{pmatrix}, \quad A \in [1\dots5].
\ee
The matrix $\Gamma_5$ is then independent from the other $\Gamma$
matrices and the coherent state formalism with six (instead of four)
operators can be used. The doubling of the Clifford space inside the
trace has to be compensated by a factor $\frac12$, such that
\eq~(\ref{WRint}) reads
\begin{equation}
\label{newWR}
W^{+}=\frac{1}{8}\int_0^{\infty}\frac{dT}{T} \Tr \exp(-T \hat{\Sigma}^2),
\end{equation} 
and the operator $\hat{\Sigma}^2$ is given by
\begin{eqnarray}
\label{Sigma2}
\hat{\Sigma}^2=(p-\Ac)^2+\Hc^2+
\frac{1}{2}\Kc_{\mu \nu} \Kc_{\mu \nu}+\frac{i}{2} \Gamma_{\mu} \Gamma_{\nu}(\Fc_{\mu \nu}+
\lbrace \Hc,\Kc_{\mu \nu}\rbrace+i\, [\Kc_{\mu \rho},\Kc_{\rho \nu}])\nonumber \\
+i\, \Gamma_{\mu} \Gamma_5(\Dc_{\mu}\Hc+\{ p_{\nu}-\Ac_{\nu},\Kc_{\mu \nu}\})-
\frac12 \Gamma_{\mu \rho \sigma} \Gamma_5 \Dc_{\mu} \Kc_{\rho \sigma}-
\frac{1}{4}\Gamma_{\mu \nu \rho \sigma} \Kc_{\mu \nu} \Kc_{\rho \sigma},
\end{eqnarray} 
with enlarged background fields defined by
\begin{equation}
\Ac_{\mu}=
\begin{pmatrix}
A_{\mu}^L & 0 \cr 0 & A_{\mu}^R \\
\end{pmatrix}, \; 
\Hc = 
\begin{pmatrix}
0 & i\, H \\ 
-i\, H^{\dagger} & 0 \\
\end{pmatrix}, 
\; \Kc_{\mu \nu} = 
\begin{pmatrix}
0 & i\,K_{\mu \nu}^s \\
 -i\,K_{\mu \nu}^{s\dagger} \\
\end{pmatrix}.
\end{equation}
$\Gamma_{A_1 ... A_k} \equiv \Gamma_{[A_1}...\Gamma_{A_k]}$ denotes
the anti-symmetrized product of $k$ $\Gamma$ matrices, and the
field-strength and the covariant derivative have been defined as
\begin{equation}
\label{defF}
\Fc_{\mu \nu}=\partial_{\mu} \Ac_{\nu}-\partial_{\nu} \Ac_{\mu}-i\, 
[\Ac_{\mu},\Ac_{\nu}], \qquad \Dc_{\mu} \Xc = \partial_{\mu} \Xc -i\, [\Ac_{\mu},\Xc].
\end{equation} 
The $\hat{\Sigma}^2$ operator is seen to be manifestly gauge and
chiral covariant. It also contains $\Gamma$ matrices to even powers
only, and is well suited for the worldline path integral
representation.

\subsubsection{Worldline Path Integral}

With the use of the coherent state formalism~\cite{coherent,Gagne1},
one can perform the transition from $\Gamma$ matrices to a path
integral over Grassman fields $\psi$, with the correspondence
$\Gamma_A\Gamma_B\rightarrow2\psi_A\psi_B$ and
$\Gamma_A\Gamma_B\Gamma_C\Gamma_D\rightarrow4\psi_A\psi_B\psi_C\psi_D$,
as long as $A$, $B$, $C$, and $D$ are all different. The final form
for the real part of the effective action is
\begin{equation}
\label{WRfinal}
W^{+}=\frac{1}{8}\int_0^{\infty}\frac{dT}{T}\Nc\int \Dc x \int_{AP} \Dc \psi\, 
\mathrm{tr}\,\Pc e^{-\int_0^Td\tau\Lc(\tau)}.
\end{equation} 
Here $\Nc$ denotes a normalization constant coming from a momentum
integration and AP stands for antiperiodic boundary conditions, which
must be fulfilled by the Grassman variables $\psi(T)=-\psi(0)$. The
Lagrangian is given by
\begin{eqnarray}
\label{LagrangianR}
\Lc(\tau)& = &\frac{\dot{x}^2}{4}+
\frac{1}{2}\psi_A\dot{\psi}_A-i\,\dot{x}_{\mu}\Ac_{\mu} + 
\Hc^2 + \frac{1}{2}\Kc_{\mu\nu}\Kc_{\mu\nu}+
2i\,\psi_{\mu}\psi_5\left(\Dc_{\mu}\Hc+
i\,\dot{x}_{\nu}\Kc_{\mu\nu}\right) \nonumber \\
&& + \, i\, \psi_{\mu}\psi_{\nu}\left(\Fc_{\mu\nu}+
\lbrace\Hc,\Kc_{\mu\nu}\rbrace  +i\, [\Kc_{\mu \rho},\Kc_{\rho \nu}] \right) \nn \\
&& - \,  \psi_{\mu}\psi_{\nu}\psi_{\rho}\left( 2 \psi_5\Dc_{\mu}\Kc_{\mu\nu}+
\psi_{\sigma}\Kc_{\mu\nu}\Kc_{\rho\sigma}\right).
\end{eqnarray}
The periodic boundary conditions for the field $x(\tau)$ suggest to
separate the zero modes of the free field operator
$\frac{d^2}{d\tau^2}$. The fields $x(\tau)$ are split into a constant
part and a $\tau$ dependent part according to $x(\tau)=x_0+y(\tau)$,
with $\partial_{\tau}x_0=0$ and $\int_0^Td\tau\,y(\tau)=0$, and the
measure in the integral is changed into $\Dc x=\Dc y\,d^Dx_0$. The
Green function is defined on a subspace orthogonal to the zero
modes. The $\psi_A$ fields contain no zero modes, so that the propagators
for the $y(\tau)$ and $\psi_A(\tau)$ fields read
\begin{eqnarray}
\label{PropR}
\langle y(\tau_1) y(\tau_2) \rangle & = & \frac{(\tau_1-\tau_2)^2}{T}-
\vert \tau_1-\tau_2 \vert, \nonumber \\
\langle \psi_{A}(\tau_1) \psi_{B}(\tau_2) \rangle & = & 
\frac{1}{2} \delta_{AB} \mathrm{sign}\left(\tau_1 - \tau_2 \right).
\end{eqnarray} 
This formalism can then be used to determine the real part of the
effective action as discussed in ref.~\cite{Gagne2}.

\subsection{Imaginary Part of the Effective Action}

As in the case of the real part of the effective action, one requires
a positive operator in order to use the Schwinger trick. Even though
this is still possible for the imaginary part, gauge and chiral
invariance cannot be manifestly conserved due to the chiral
anomaly. For example, in ref.~\cite{Gagne1,Gagne2} a parameter
$\alpha$ is introduced, which breaks the chiral invariance, but leads
to a positive operator. However the resulting expression is not
appropriate for higher order calculations since the breaking of
manifest chiral invariance leads to a large number of contributions in
the perturbative expansion of the path integral.

The aim of the present work is to present a worldline representation
of the effective current for which a manifestly chiral covariant
expression exists. This current can then be integrated to obtain the
effective action~\cite{Hoker,Salcedo2,Konstandin}.  This integration
rather proceeds by matching: First, a general effective action is
proposed, which has the expected chiral and covariant properties. The
functional variation of this action is then matched to the covariant
current that is obtained using the worldline formalism.  This method
has the advantage that it is both gauge and chiral invariant at each
stage of the calculation. The anomaly only leads to additional
complications in the matching procedure of the lowest order
contributions as will be discussed in detail in the next section.

Starting point of our analysis is the functional derivative of the
imaginary part of the effective action in \eq~(\ref{Sep})
\begin{equation}
\label{WI}
\delta W^{-}=\frac{1}{2}\delta\left(\log\Det \Oc-
\log\Det \Oc^{\dagger}\right)=\frac{1}{2}\Tr\left(\delta\Oc\frac{1}{\Oc}-
\delta\Oc^{\dagger}\frac{1}{\Oc^{\dagger}}\right).
\end{equation} 
This expression can be rewritten in terms of a positive operator which
can be used to employ the worldline representation in combination with
the heat kernel formula. Incidentally, it can also be expressed in a
manifestly chiral covariant form, what simplifies higher order
calculations tremendously as compared to the formalism presented in
ref.~\cite{Gagne2}.

\subsubsection{Construction of a Positive Operator for 
the Imaginary Part of the Effective Action}

The expression in \eq~(\ref{WI}) can be transformed using the
operator $\Sigma$ defined in \eq~(\ref{WR})
\begin{equation}
\delta W^{-}=\frac{1}{2}\Tr 
\begin{pmatrix} 
0 & \delta\Oc \\ 
-\delta\Ocd & 0 \\
\end{pmatrix} 
\begin{pmatrix} 
0 & 1/\Ocd \\ 
1/\Oc & 0 \\
\end{pmatrix},
\end{equation}
which, with the introduction of a new matrix $\chi$, can be rewritten
as
\begin{equation}
\label{dWI}
\delta W^{-}=\frac{1}{2}\Tr \chi \delta\Sigma \Sigma^{-1}, 
\end{equation} 
with
\begin{equation}
\Sigma=
\begin{pmatrix}
0 & \Oc \\
 \Oc^{\dagger} & 0 \\
\end{pmatrix}, \qquad 
\chi = 
\begin{pmatrix}
\Eins_4 & 0 \\ 
0 & -\Eins_4 \\
\end{pmatrix}.
\end{equation}
To produce the positive definite operator $\Sigma^2$ in
\eq~(\ref{dWI}), we multiply and divide by $\Sigma$, using the cyclic
property of the trace and the fact that $\Sigma$ anticommutes with
$\chi$, to obtain
\begin{eqnarray}
\label{dWIplusb}
\delta W^{-}&=&\frac{1}{4}\Tr \left( \chi\delta\Sigma\Sigma + 
\Sigma\chi\delta\Sigma\right) \Sigma^{-2}
\nonumber\\ 
&=&\frac{1}{4}\Tr\chi\left[\delta\Sigma,\Sigma\right]\Sigma^{-2}.
\end{eqnarray} 
Since the last factor is a positive operator, it can be reexpressed as
an integral, similar to the expression of the real part of the
effective action in \eq~(\ref{Sigma2}), namely
\begin{eqnarray}
\label{dWIplus}
\delta W^{-}&=&\frac{1}{4}\Tr\int_0^{\infty}dT
\chi\left[\delta\Sigma,\Sigma\right]e^{-T\Sigma^2}.
\end{eqnarray}
As in the case for the real part, the chiral covariance can be made
manifest by changing to an appropriate basis. With the help of the
matrix $M$ in \eq~(\ref{Change}), one obtains again
\begin{equation}
\label{newSigma2}
\tilde{\Sigma}= \gamma_{\mu}(p_{\mu}-\Ac_{\mu})-\gamma_5 \Hc-
\frac{i}{2} \gamma_{\mu} \gamma_{\nu} \gamma_5 \Kc_{\mu \nu}.
\end{equation} 
The additional factors $\chi \left[\delta\Sigma,\Sigma\right]$ read
\begin{equation}
\label{chitilde}
M^{-1} \chi M = \tilde{\chi} = 
\begin{pmatrix}
\gamma_5 & 0 \\ 
0 & -\gamma_5 \\
\end{pmatrix}
= \chi \,\gamma_5,
\end{equation}
and for the case $\delta\tilde\Sigma=-\gamma_{\mu}\delta\Ac_{\mu}$
\begin{eqnarray}
\label{commutator}
\left[ \delta\tilde\Sigma ,\tilde\Sigma \right]&=&-
\gamma_{\mu\nu}\left\lbrace \delta\Ac_{\mu},p_{\nu}-
\Ac_{\nu}\right\rbrace-i\,\Dc_{\mu}\delta\Ac_{\mu}-
\gamma_5\gamma_{\mu}\left\lbrace\delta\Ac_{\mu},\Hc\right\rbrace\nonumber\\
&&+i\,\gamma_5\gamma_{\mu}\left[\delta\Ac_{\nu},\Kc_{\mu\nu}\right]-
\frac{i}2\,\gamma_5\gamma_{\mu\lambda\sigma}
\left\lbrace\delta\Ac_{\mu},\Kc_{\lambda\sigma}\right\rbrace.
\end{eqnarray} 

To use the coherent state formalism, it is again necessary to enlarge
the Clifford algebra and to replace the $\gamma$ matrices by $\Gamma$
matrices. However, taking into account the factor $\gamma_5$ in
\eq~(\ref{chitilde}) the imaginary part of the effective action contains only
odd combinations of $\gamma$ matrices. Thus, the replacement
\be
\gamma_A \to \Gamma_A = \gamma_A \otimes
\begin{pmatrix}
0 & 1 \\ 
1 & 0 \\
\end{pmatrix}, \quad A \in [1\dots5]
\ee
has to be compensated by a factor
\begin{equation}
\label{compensator}
-\frac{i}2 \Gamma_7\Gamma_6=  \Eins_4 \otimes
\begin{pmatrix}
0 & \frac12 \\ 
\frac12 & 0 \\
\end{pmatrix}.
\end{equation}
The overall factor $\Gamma_7$ changes the boundary condition of
the fermionic sector from antiperiodic to periodic as explained in
ref.~\cite{Gagne1}. This means that the fermionic sector contains
zero modes, which have to be separated in the same way as was done
for the bosonic sector.

Including the factor in \eq~(\ref{compensator}) to compensate for the
doubling of the Clifford space, one obtains
\begin{equation}
\label{WIm2}
\delta W^{-}=\frac{i}{8}\Tr\int_0^{\infty}dT\Gamma_7\Gamma_6\chi w(T)e^{-T\hat\Sigma^2},
\end{equation} 
where $\hat\Sigma^2$ is given in \eq~(\ref{Sigma2}), and the insertion
due to the commutator yields
\begin{eqnarray}
w(T)&=&-\frac{1}{2}\Gamma_5\Gamma_{\mu\nu}\left\lbrace \delta\Ac_{\mu},p_{\nu}-
\Ac_{\nu}\right\rbrace-i\,\Gamma_5\Dc_{\mu}\delta\Ac_{\mu}-
\Gamma_{\mu}\left\lbrace\delta\Ac_{\mu},\Hc\right\rbrace \nonumber\\
&&+i\,\Gamma_{\mu}\left[\delta\Ac_{\nu},\Kc_{\mu\nu}\right]-
\frac{i}{2} \Gamma_{\mu\lambda\sigma}
\left\lbrace\delta\Ac_{\mu},\Kc_{\lambda\sigma}\right\rbrace.
\end{eqnarray} 
To transform this expression into a worldline path integral, a similar
procedure as for the real part of the effective action can be
followed. Products of $\Gamma$ matrices can be replaced by Grassman
fields, however in this case the Jacobian of the transformation
contains additional contributions from the zero modes
\begin{eqnarray}
\Dc\theta\Dc\bar\theta &\equiv& 
d\theta_3d\theta_2d\theta_1d\bar\theta_1d\bar\theta_2d\bar\theta_3
\Dc\theta'\Dc\bar\theta' \nonumber\\
&=&\frac{1}{J}d\psi_1^0d\psi_2^0d\psi_3^0d\psi_4^0d\psi_5^0d\psi_6^0\Dc\psi'.
\end{eqnarray} 
The factor $J$ only includes the Jacobian for the zero modes, while
the Jacobian for the orthogonal modes is absorbed in the normalization
of the correlation functions of the $\psi_A'$. $J$ can be calculated
from the definition of the Grassman fields $\psi$ in the coherent
state formalism~\cite{Gagne1} and yields in $D$ dimension
\begin{equation}
J=\det\left(\frac{\partial\theta,\bar\theta}{\partial\psi}\right)=(-i)^{(D+2)/2}.
\end{equation} 

The final result can be expressed as
\begin{equation}
\label{current}
\delta W^{-}=\frac{1}{8} \, \tr \int_0^{\infty}dT \Nc\int\Dc x 
\int_{P}\Dc\psi \,\chi w(T)\Pc e^{-\int_0^Td\tau\Lc(\tau)}. 
\end{equation} 
The Lagrangian is of the same form as in the real
part, \eq~(\ref{LagrangianR}),
\begin{eqnarray}
\label{LagrangianI}
\Lc(\tau)& = &\frac{\dot{x}^2}{4}+\frac{1}{2}\psi_A\dot{\psi}_A-
i\,\dot{x}_{\mu}\Ac_{\mu} + 
\Hc^2-\frac{1}{2}\Kc_{\mu\nu}\Kc_{\mu\nu}+2i\,\psi_{\mu}\psi_5\left(\Dc_{\mu}\Hc+
i\,\dot{x}_{\nu}\Kc_{\mu\nu}\right) \nonumber \\
& + & i\, \psi_{\mu}\psi_{\nu}\left(\Fc_{\mu\nu}+
\lbrace\Hc,\Kc_{\mu\nu}\rbrace\right) - 
2\psi_{\mu}\psi_{\nu}\psi_{\rho}\left(\psi_5\Dc_{\mu}\Kc_{\mu\nu}+
\frac{1}{2}\psi_{\sigma}\Kc_{\mu\nu}\Kc_{\rho\sigma}\right).
\end{eqnarray}
and the trivial integration over $\psi_6$ can been carried
out, so that the insertion yields
\begin{eqnarray}
\label{InsertionI}
w(T)&=&-4i\,\psi_5\psi_{\mu}\psi_{\nu}\delta\Ac_{\mu}\dot{x}_{\nu}-
2i\,\psi_5\Dc_{\mu}\delta\Ac_{\mu}-
2\psi_{\mu}\left\lbrace\delta\Ac_{\mu},\Hc\right\rbrace\nonumber\\
&&+2i\,\psi_{\mu}\left[\delta\Ac_{\nu},\Kc_{\mu\nu}\right]-
2i\,\psi_{\mu}\psi_{\lambda}\psi_{\sigma}
\left\lbrace\delta\Ac_{\mu},\Kc_{\lambda\sigma}\right\rbrace.
\end{eqnarray} 
The normalization $\Nc$ coming from the momentum integration, satisfies
\begin{equation}
\Nc\int\Dc x e^{-\int_0^Td\tau\frac{\dot{x}}{4}}=(4\pi T)^{-D/2}\int d^D x.
\end{equation} 
The Green function for the bosonic field $x$ is the same as for the
real part of the effective action, \eq~(\ref{PropR}), while the Green
function of the Grassman fields $\psi_A$ differs due to the presence
of the zero modes. The fermionic fields are split according to
$\psi_A(\tau)=\psi_A^0+\psi_A^{'}(\tau)$, with
$\partial_{\tau}\psi_A^0=0$ and $\int_0^Td\tau\psi_A^{'}(\tau)=0$ and
the measure turns into
$\Dc\psi=d\psi_1d\psi_2d\psi_3d\psi_4d\psi_5 \Dc\psi'$. The Green
function for the $\psi_A^{'}$ fields, defined on a space orthogonal to
the zero modes, reads
\begin{equation}
\left\langle \psi_A^{'}(\tau_1)\psi_B^{'}(\tau_2) \right\rangle = 
\delta_{A B}\left( \frac{1}{2} \textrm{sign}(\tau_1-\tau_2)-\frac{(\tau_1-\tau_2)}{T}\right).
\end{equation}  
These results can be easily generalized to different dimensions. In
two dimension, one obtains an additional overall factor $-i$ from the
Jacobian of the zero modes and the fermionic measure reads
$\Dc\psi=d\psi_1d\psi_2d\psi_5 \Dc\psi'$.

\subsubsection{The Effective Density}

The effective density is obtained by varying with respect to the $\Hc$
field, so that $\delta\tilde\Sigma=-\gamma_5\delta\Hc$. In comparison
to the worldline representation of the covariant current only the
insertion changes into
\begin{equation}
\label{commDensity}
\left[ \delta\tilde\Sigma ,\tilde\Sigma \right]=
-\gamma_5\gamma_{\mu}\left\lbrace \delta\Hc,p_{\mu}-
\Ac_{\mu}\right\rbrace+\left[\delta\Hc,\Hc\right]+
\frac{i}{2}\gamma_{\mu}\gamma_{\nu}\left[\delta\Hc,\Kc_{\mu\nu}\right].
\end{equation} 
The corresponding insertion $w(T)$ in the path integral reads then
\begin{equation}
\label{insertionDensity}
w(T)=-2i\,\psi_{\mu}\dot{x}_{\mu}\delta\Hc+
2\psi_5\left[\delta\Hc,\Hc\right]+
2i\,\psi_{\mu}\psi_{\nu}\left[\delta\Hc,\Kc_{\mu\nu}\right].
\end{equation} 
Since $\delta\Ac$ carries an index, the effective current is of one
order lower than the effective density and usually results in less
terms to calculate. The advantage of the effective density lies in
the matching process, since the factors in the effective density
consist of the same type as found in the effective action. They both
combine the same type of object, $\Dc \Hc$ and $\Fc$, to the same kind of
order, while the effective current combines the terms to a lower
order. Besides, there is no distinction between a consistent effective
density and a covariant effective density, as there is for the
effective current, as will be explained in the next section.

\subsubsection{Distinction between the Consistent and the Covariant Current}

With \eq~(\ref{current}) an expression for the covariant current which
is chiral and gauge covariant was derived. This current cannot be
the variation of the effective action, since the effective action
contains the chiral anomaly, and in fact the covariant current is not
a variation of any action. The reason for this is that performing the
variation does not commute with the regularization procedure we used,
namely the Schwinger trick. On the other hand, knowing the chiral
anomaly, one can reproduce the so-called consistent current that
denotes the true variation of the effective action.

To explain the relation between the two currents, we define a general
variation
\begin{equation}
\delta_Y = \int dx \, Y_{\mu}^a(x) \frac{\delta}{\delta \Ac_{\mu}(x)},
\end{equation} 
so that a gauge variation $\delta_{\xi}$ is given by
\begin{equation}
\label{GaugeVar}
\delta_{\xi}=\int dx \, \left(\Dc_{\mu}\xi\right)(x)\frac{\delta}{\delta \Ac_{\mu}(x)}.
\end{equation} 
Two subsequent variations have then the commutator
$[\delta_{Y},\delta_{\xi}]=\delta_{[Y,\xi]}$ and in order to find the
transformation properties of the consistent current, one can apply
this commutator to the effective action
\begin{equation}
\label{GaugeComm}
[\delta_Y,\delta_{\xi}]W^{-}[\Ac_{\mu}]=
\delta_{[Y,\xi]}W^{-}[\Ac_{\mu}].
\end{equation} 
Using the anomalous Ward identity~\cite{Anomaly}
\begin{equation}
\label{AnomWard}
\delta_{\xi}W^{-}[\Ac_{\mu}]=\int dx \, \xi(x)G[\Ac_{\mu}](x),
\end{equation} 
with $G[\Ac_{\mu}]$ denoting the consistent anomaly, one can
evaluate both sides of \eq~(\ref{GaugeComm}) to obtain
\bea
\int dx \, [Y_{\mu},\xi](x)\frac{\delta}{\delta \Ac_{\mu}(x)}W^{-}[\Ac_{\mu}] &=&
\delta_Y\int dx \, \xi(x)G[\Ac_{\mu}](x) \nn \\ 
&& -\delta_{\xi}\int dx \, Y_{\mu}(x)\frac{\delta}{\delta \Ac_{\mu}(x)}W^{-}[\Ac_{\mu}].
\eea
Defining the consistent current as the variation of the effective
action
\begin{equation}
\langle j^{\mu}(x)\rangle=\frac{\delta}{\delta \Ac_{\mu}(x)}W^{-}[\Ac_{\mu}],
\end{equation} 
one finds
\begin{eqnarray}
\int dx \, Y_{\mu}(x)\delta_{\xi}\langle j^{\mu}(x)\rangle =
\int dx \, Y_{\mu}[\langle j^{\mu}(x)\rangle,\xi](x)+
\int dx \, \xi(x)\delta_YG[\Ac_{\mu}](x).
\end{eqnarray} 
Since $Y$ was a general variation this leads to
\begin{equation}
\delta_{\xi}\langle j^{\mu}(x)\rangle=[\langle j^{\mu}(x)\rangle,\xi]+
\int dy \, \xi^b(y)\frac{\delta}{\delta \Ac_{\mu}(x)}G[\Ac_{\mu}](y).
\end{equation} 
This shows that only if the anomaly vanishes, the current transforms
covariantly. This relation can be used to determine the connection
between the consistent current, i.e. the true variation of the action,
and the covariant current. The latter is obtained by adding an object
$P^{\mu}[\Ac_{\mu}]$, called the Bardeen-Zumino polynomial
\cite{BardeenZumino}, to the consistent current so that the sum transforms
covariantly
\begin{equation}
\label{defCovCurrent}
\langle\bar{j^{\mu}}\rangle=\langle j^{\mu}\rangle+\langle P^{\mu}\rangle.
\end{equation} 
This implies the following gauge transformation property for the BZ polynomial 
\begin{equation}
\label{condP}
\delta_{\xi}P^{\mu}[\Ac_{\mu}](x)=
\left[P^{\mu}[\Ac_{\mu}],\xi\right](x)- 
\int dy \, \xi(y)\frac{\delta}{\delta \Ac_{\mu}(x)}G[\Ac_{\mu}](y).
\end{equation} 
It is not obvious that such an object exists, but using
\begin{equation}
\label{defP}
P^{\mu}[\Ac_{\mu}]=\frac{1}{48\pi^2}\epsilon^{\mu\nu\lambda\sigma} 
\tr \chi\left(\Ac_{\nu}\Fc_{\lambda\sigma}+
\Fc_{\lambda\sigma}\Ac_{\nu} + i\,\Ac_{\nu}\Ac_{\lambda}\Ac_{\sigma}\right),
\end{equation} 
and the consistent anomaly~\cite{Anomaly}
\begin{equation}
G[\Ac_{\mu}]=\frac{1}{24\pi^2}\epsilon^{\mu\nu\lambda\sigma} 
\tr \, \chi \, \partial_{\mu}\left(\Ac_{\nu}\partial_{\lambda}\Ac_{\sigma} - 
\frac{i}{2}\Ac_{\nu}\Ac_{\lambda}\Ac_{\sigma}\right),
\end{equation} 
it can be shown that the definition of $P^{\mu}$ in \eq~(\ref{defP})
provides a unique polynomial in $\Ac_{\mu}$ that satisfies
\eq~(\ref{condP}). The corresponding functions in two dimensions are
given by
\be
P^{\mu}= \frac{1}{4\pi}\epsilon^{\mu\nu}\tr \Ac_{\nu}, \quad
G[\Ac_{\mu}]=\frac{1}{4\pi}\epsilon^{\mu\nu} 
\tr \, \chi \, \partial_{\mu} \, \Ac_{\nu}.
\ee

As stated above, the path integral in \eq~(\ref{current}) constitutes
a worldline representation of the covariant current. To obtain the
imaginary part of the effective action from the covariant current one
can use the following ansatz
\begin{equation}
\label{ansatz_action}
W^{-}=\Gamma_{gWZW}+W_c^{-}.
\end{equation} 
Here, $\Gamma_{gWZW}$ is an extended gauged Wess-Zumino-Witten
action~\cite{WZ,Witten,Hoker}, which is chosen to reproduce
the correct chiral anomaly, and $W_c^{-}$ denotes a chiral invariant
part. The variation of the functional $\Gamma_{gWZW}$, consists of
a part that saturates the anomaly, namely the BZ polynomial, and a
covariant remainder which has to be added to the variation of $W_c^{-}$
to yield the covariant current.

\subsubsection{The Wess-Zumino-Witten action}

When the effective action is separated into two parts, it is required
by the non-covariant part that it reproduces the anomaly. It is well
known that the WZW action has this property. 

The ungauged WZW action in four dimension is e.g. of the form

\begin{equation}
\label{ungWZW}
\Gamma(\Ub)=\frac{i}{48\pi^2}\int_Qd^5x \, \epsilon^{abcde}
\tr \left[\frac{1}{5}\Ub^{-1}\partial_a\Ub\Ub^{-1}\partial_b\Ub
\Ub^{-1}\partial_c\Ub\Ub^{-1}\partial_d\Ub\Ub^{-1}\partial_e\Ub\right],
\end{equation} 
where Q is a five-dimensional space with boundary $\partial Q$ equal
to the $R^4$ flat Euclidean space. The matrix $\Ub$ is a unitary
matrix, and is usually related to the case where the mass can be
expressed as a constant times that unitary matrix. We are interested
in the more general case when the mass matrix is not of this form
which is called extended WZW action. In addition, the presence of the
background gauge fields makes a gauging of the action mandatory. The
gauged extended WZW action can be generally expressed as the integral
in five dimensions~\cite{Hoker}. Unlike the action itself, the
resulting current turns out to be a total derivative in five
dimensions, such that it can be represented by an integral over the
physical four-dimensional space
\begin{eqnarray}
\delta\Gamma_{gWZW}&=&
\frac{1}{96\pi^2}\int d^4x \, \epsilon^{\mu\nu\lambda\sigma}\tr \,\chi 
\left[\delta\Ac_{\mu}\left(
-\Hc^{-1}\Dc_{\nu}\Hc\Hc^{-1}\Dc_{\lambda}\Hc\Hc^{-1}\Dc_{\sigma}\Hc\right.\right.\nonumber\\
&&\quad+\Dc_{\nu}\Hc\Hc^{-1}\Dc_{\lambda}\Hc\Hc^{-1}\Dc_{\sigma}\Hc\Hc^{-1}-
i\,\left\lbrace\Hc^{-1}\Dc_{\nu}\Hc-\Dc_{\nu}\Hc\Hc^{-1},\Fc_{\lambda\sigma}\right\rbrace\nonumber\\
&&\quad+\frac{i}{2}\Hc\left\lbrace\Hc^{-1}\Dc_{\nu}\Hc,\Fc_{\lambda\sigma}\right\rbrace\Hc^{-1}
-\frac{i}{2}\Hc^{-1}\left\lbrace\Dc_{\nu}\Hc\Hc^{-1},\Fc_{\lambda\sigma}\right\rbrace\Hc\nonumber\\
&&\quad\left.\left.-2\left\lbrace\Ac_{\nu},\Fc_{\lambda\sigma}\right\rbrace-
2i\,\Ac_{\nu}\Ac_{\lambda}\Ac_{\sigma}\right)\right],
\end{eqnarray} 
or in two dimensions
\begin{eqnarray}
\label{Contr2d}
\delta\Gamma_{gWZW}&=&
\frac{1}{8\pi}\int d^2x \,\epsilon^{\mu\nu} \tr \, \chi\left[\delta\Ac_{\mu}
\left(-i\,\Hc^{-1}\Dc_{\nu}\Hc+i\,\Dc_{\nu}\Hc\Hc^{-1}-2\Ac_{\nu}\right)\right].
\end{eqnarray} 
Notice that in both cases the last term in the current denotes the BZ
polynomial. The remaining chiral covariant terms have to be subtracted
from the covariant current before it is matched to the effective
action according to the ansatz made in \eq~(\ref{ansatz_action}).

\section{Lowest Order Effective Action\label{sec_low}}

\subsection{Effective covariant current}

In order to reproduce the results from ref.~\cite{Salcedo2}, we
neglect in this section the antisymmetric field $K_{\mu\nu}$. The
fields $\Ac$ and $\Hc$ are matrices of some internal group, and we
only assume that $\Hc(x_0)$ is nowhere singular. With this in mind, we
restate our result \eq~(\ref{current}) from the last section in $D$
dimensions
\begin{equation}
\label{dWIm2}
\delta W^{-}=-\frac{i^{D/2}}{8}\, \mathrm{tr}\int_0^{\infty}dT \Nc
\int\Dc x \int_{P}\Dc\psi \,\chi w(T)\Pc e^{-\int_0^Td\tau\Lc(\tau)},
\end{equation}
with
\begin{eqnarray}
\label{ohneK}
\Lc(\tau) &=&\frac{\dot{x}^2}{4}+\frac{1}{2}\psi_A\dot{\psi}_A-
i\,\dot{x}_{\mu}\Ac_{\mu} + \Hc^2+2i\,\psi_{\mu}\psi_5\Dc_{\mu}\Hc+
i\,\psi_{\mu}\psi_{\nu}\Fc_{\mu\nu},\nonumber\\
w(T)&=&-4i\,\psi_5\psi_{\mu}\psi_{\nu}\delta\Ac_{\mu}\dot{x}_{\nu}-
2i\,\psi_5\Dc_{\mu}\delta\Ac_{\mu}-2\psi_{\mu}\left\lbrace\delta\Ac_{\mu},\Hc\right\rbrace.
\end{eqnarray} 
Next, the derivative expansion of the heat kernel is used. In the
derivative expansion terms are classified by the number of covariant
indices that they carry, so that $\Dc_{\mu}\Hc$ is of first order,
while $\Fc_{\mu\nu}$ is of second order. The worldline formalism is
well suited for this expansion, and there are two major advantages
compared to the more traditional methods used e.g. in
ref.~\cite{Salcedo2}. First, the tedious manipulations using the
$\gamma$ algebra are avoided. Secondly, the momentum integration is
omitted and replaced by the rather trivial integration in $\tau$
space.

The coordinate is split as $x(\tau)=x_0+y(\tau)$, and we work in the
Fock-Schwinger gauge~\cite{Fockgauge2}, in which $\Ac(x)\cdot y=0$. In
this gauge, expressions remain gauge covariant and the field $\Ac$ can
be expressed in terms of the field strength tensor $\Fc_{\mu\nu}$ by
\begin{equation}
\label{expA}
\Ac_{\mu}(x)=\int_0^1d\alpha \, \alpha \, \Fc_{\rho\mu}(x_0+\alpha y)y_{\rho}.
\end{equation} 
All background fields can then be expanded around the point $x_0$ in terms
of covariant derivatives
\begin{equation}
X(x_0+y(\tau))=\exp\left(y(\tau)\cdot\Dc_{x_0}\right)X(x_0),
\end{equation} 
where $\Dc_{x_0}$ refers to the covariant derivative in
\eq~(\ref{defF}) with respect to $x_0$. With the expansion of the
field strength tensor in terms of covariant derivatives and
\eq~(\ref{expA}), one can rewrite the field $\Ac$ as
\begin{eqnarray}
\Ac_{\mu}(x)= \frac{1}{2}y_{\rho}\Fc_{\rho\mu}(x_0)+
\frac{1}{3}y_{\alpha}y_{\rho}\Dc_{\alpha}\Fc_{\rho\mu}(x_0)
+\frac{1}{4\cdot2!}y_{\alpha}y_{\beta}y_{\rho}
\Dc_{\alpha}\Dc_{\beta}\Fc_{\rho\mu}(x_0)+\ldots \,\,.
\end{eqnarray} 
 
Since we will not carry out the integration with respect to $x_0$ we
use the following notation in $D$ dimensions
\begin{equation}
\label{norm}
\left\langle X\right\rangle_D= - \left( \frac{i}{4\pi} \right)^{D/2}
\mathrm{tr}\chi\int d^D x_0 X.
\end{equation} 
It is important to remember that $\chi$ and $\Hc$ anticommute; hence, when
the cyclic property of the trace is used, a minus sign is
generated, for example
\bea
\label{norm_ex}
\left\langle \epsilon^{\mu\nu\lambda\sigma}
\Hc \Fc_{\mu\nu} \Hc^3 \Fc_{\lambda\sigma}\right\rangle&=&
-\left\langle \epsilon^{\mu\nu\lambda\sigma} 
\Fc_{\mu\nu} \Hc^3 \Fc_{\lambda\sigma}\Hc\right\rangle=
-\left\langle \epsilon^{\mu\nu\lambda\sigma} 
\Hc^3 \Fc_{\lambda\sigma}\Hc \Fc_{\mu\nu} \right\rangle \nn \\
&=&-\left\langle \epsilon^{\mu\nu\lambda\sigma} 
\Hc^3 \Fc_{\mu\nu}\Hc \Fc_{\lambda\sigma} \right\rangle.
\eea
After expanding the mass field
$\Hc(x)^2=\Hc^2(x_0)+y_{\mu}\Dc_{\mu}\Hc^2(x_0)+\ldots$, the field
$\Hc(x_0)$ is treated non-perturbatively. Since all the fields can be
matrices of some internal space the resulting expressions normally
cannot be expressed in closed form.  For this case we use the labeled
operator notation laid down in ref.~\cite{Feynman2,Salcedo1}. The
notation works as follows: In an expression
$f(A_1,B_2,\ldots)XY\ldots$, the labels of the operators $A$, $B$,
$\ldots$ denote the position of that operator with respect to the
remaining operators $XY\ldots$. For instance, for the function
$f(A,B)=\alpha(A)\beta(B)$, the expression $f(A_1,B_2)XY$ represents
$\alpha(A)X\beta(B)Y$. In the case at hand, the operator appearing in
the functions is always $m:=\Hc(x_0)$, such that general functions $f$
can be easily interpreted in the basis where $m$ is diagonal. Using
this notation, \eq~(\ref{norm_ex}) can be recast as
\bea
\left\langle \epsilon^{\mu\nu\lambda\sigma}
\Hc \Fc_{\mu\nu} \Hc^3 \Fc_{\lambda\sigma}\right\rangle&=&
\left\langle \epsilon^{\mu\nu\lambda\sigma} 
m_1 m_2^3 \Fc_{\mu\nu}\Fc_{\lambda\sigma}\right\rangle=
-\left\langle \epsilon^{\mu\nu\lambda\sigma} 
m_3 m_2^3 \Fc_{\mu\nu}\Fc_{\lambda\sigma}\right\rangle \nn \\
&=& -\left\langle \epsilon^{\mu\nu\lambda\sigma} 
m_2 m_1^3 \Fc_{\mu\nu}\Fc_{\lambda\sigma}\right\rangle
=-\left\langle \epsilon^{\mu\nu\lambda\sigma} 
\Hc^3 \Fc_{\mu\nu}\Hc \Fc_{\lambda\sigma} \right\rangle.
\eea
This notation can also be used to simplify the matrix valued
derivative. Using the definition 
\begin{equation}
(\nabla f )(m_1,m_2) := \frac{f(m_1)-f(m_2)}{m_1-m_2},
\end{equation} 
it is possible to prove that 
\be
\Dc_{\mu}f(m)=(\nabla f)(m_1,m_2) \, \Dc_{\mu} \Hc.
\ee
For example, in the polynomial case $f(m)=m^3$ one obtains
\bea
\Dc_{\mu}f(m) &=& \Dc_{\mu} (\Hc^3) 
= \Dc_{\mu} \Hc \, \Hc^2 + \Hc \Dc_{\mu} \Hc \, \Hc + \Hc^2 \Dc_{\mu} \Hc \nn \\
&=& (m_2^2 + m_1 m_2 + m_1^2 ) \, \Dc_{\mu} \Hc 
= \frac{m_1^3 - m_2^3}{m_1 - m_2} \, \Dc_{\mu} \Hc \nn \\
&=& (\nabla f)(m_1,m_2) \, \Dc_{\mu} \Hc.
\eea
As mentioned earlier, non-polynomial expressions are hereby
interpreted in a basis where $m$ is diagonal, so that for
$m=\textrm{diag}(d_1, \dots, d_n)$
\be
\frac{f(m_1)-f(m_2)}{m_1-m_2} \, X 
= \frac{f(d_i)-f(d_j)}{d_i-d_j} \, X_{ij}. 
\ee

More general, this suggests the following definition for the case with several
variables:
\begin{equation}
\nabla_k f(m_1,\ldots,m_n) = 
\frac{f\left(m_1,\ldots,{\hat m}_{k+1},\ldots,m_n\right)
-f\left(m_1,\ldots,\hat{m_{k}},\ldots,m_n\right)}{m_k-m_{k+1}},
\end{equation} 
where $\hat{m_k}$ indicates that the corresponding argument is left
out. 

If all arguments of the functions are of the same type one can further
simplify the notation and use subscripts to refer to the argument of
the function, e.g.  $f(m_1,m_2)=:f_{12}$ and we employ this notation
in the following. Additionally, negative arguments will be denoted by
underlining the corresponding index,
$f(-m_1,m_2)=:f_{\underline{1}2}$. More applications of the labeled
operator notation can be found in refs.~\cite{Salcedo1,Salcedo2}.

The path ordering in \eq~(\ref{dWIm2}) is defined by
\begin{equation}
\label{defPath}
\Pc\prod_{i=1}^N\int_0^Td\tau_i\equiv N!
\int_0^Td\tau_1\int_0^{\tau_1}d\tau_2\cdots\int_0^{\tau_{N-1}}d\tau_N=
N!\int_0^Td\tau_1\cdots\int_0^Td\tau_N\prod_{i-1}^{N-1}\theta(\tau_i-\tau_{i+1}).
\end{equation} 
Separating the Lagrangian \eq~(\ref{ohneK}) into
$\Lc(\tau)=\Lc_0(\tau)+\Hc^2(x_0)+\Lc_1(\tau)$, with
\begin{eqnarray}
\Lc_0(\tau)&=&\frac{\dot{x}^2}{4}+\frac{1}{2}\psi_A\dot{\psi}_A, \nonumber\\
\Lc_1(\tau)&=&-i\,\dot{x}_{\mu}\Ac_{\mu}(x)+2i\,\psi_{\mu}\psi_5\Dc_{\mu}\Hc(x)+
i\,\psi_{\mu}\psi_{\nu}\Fc_{\mu\nu}(x)+y_{\mu}\Dc_{\mu}\Hc^2(x_0)+\ldots,
\end{eqnarray} 
the terms of the expansion of $\Hc^2(x)$, except the leading term
$\Hc^2(x_0)$, are attributed to $\Lc_1(\tau)$, and treated
perturbatively. Notice that $\Lc_0$ commutes with the rest of the
Lagrangian, so that the expansion of the path ordered exponential in
\eq~(\ref{dWIm2}) takes the form
\begin{eqnarray}
\label{expansionPath}
\Pc e^{-\int_0^T d\tau \Lc(\tau)}=e^{-\int_0^T d\tau \Lc_0(\tau)}
\left(e^{-T\Hc^2(x_0)}+
\int_0^Td\tau_1e^{-(T-\tau_1)\Hc^2(x_0)}\left(-\Lc_1(\tau_1)\right)
e^{-\tau_1\Hc^2(x_0)}\right.\nonumber\\
\quad\left.
+\int_0^Td\tau_1\int_0^{\tau_1}d\tau_2e^{-(T-\tau_1)\Hc^2(x_0)}
\left(-\Lc_1(\tau_1)\right)e^{-(\tau_1-\tau_2)\Hc^2(x_0)}\left(-\Lc_1(\tau_2)\right)
e^{-\tau_2\Hc^2(x_0)}+\ldots\right).
\end{eqnarray} 
When performing the $\psi$ integrals, the zero modes have to be
saturated and at least a factor $\psi_1^0 \dots\psi_D^0\psi_5^0$ is
required from the Grassman fields in order to contribute. The first
term in
\eq~(\ref{expansionPath}) lacks the appropriate $\psi$ factor except
in two dimensions, where the first term of the insertion
\eq~(\ref{ohneK}) already has the appropriate factor. However it
contains a factor $\dot{x}_{\mu}$ which must be contracted with a
similar factor to form a Green function, hence it does not contribute
and can be left out. The rest of \eq~(\ref{expansionPath}) can be
simplified using the labeled operator notation. Using the expression
$m_n^2$ to denote $\Hc^2(x_0)$ in $n$th position, one obtains
\begin{eqnarray}
\label{expansionPath2}
\Pc e^{-\int_0^T d\tau \Lc(\tau)}&=&e^{-\int_0^T d\tau \Lc_0(\tau)}\left(
-\int_0^Td\tau_1e^{-T m_1^2-\tau_1\left(m_2^2-m_1^2\right)}\Lc_1(\tau_1)
\right.\nonumber\\
&&\hskip -1 cm \left.+\int_0^Td\tau_1\int_0^{\tau_1}d\tau_2e^{-T m_1^2-
\tau_1\left(m_2^2-m_1^2\right)-
\tau_2\left(m_3^2-m_2^2\right)}\Lc_1(\tau_1)\Lc_1(\tau_2)+\ldots\right).
\end{eqnarray}
The evaluation of the worldline path integral can be summarized as
follows: First, all fields in \eq~(\ref{expansionPath2}) and the
insertion are expanded around $x_0$. Next, the functional integration
over the $y$ fields is carried out, generating bosonic Green
functions. Then, the $\psi$ integrations are performed saturating the
zero modes and generating fermionic Green functions. Finally, the $T$
and $\tau$ integrations are performed.

Before presenting the actual calculation, we comment on the behavior
of the effective action under complex conjugation. As noted earlier,
in any contribution to the imaginary part of the action the field
$\psi_5$ appears an odd number of times. If one attributes a factor
$i$ to the operators $\Fc$ and $\delta\Ac$ one observes that the
remaining expressions in the current in \eq~(\ref{ohneK}) are
real. Accordingly, all expressions in $W^-$ are real as long as a
factor $i$ is attributed to the operator $\Fc$. In addition, notice
that the effective action has to be an even function in the masses due
to chiral invariance.

In order to showcase the method, we present the lowest order
calculation in two dimensions. The lowest order contribution coming
from the first term in the insertion is given by
\begin{equation}
-4i\,\psi_5\psi_{\mu}\psi_{\nu}\dot{y}_{\nu}(T)
\int_0^Td\tau_1e^{-T m_1^2-\tau_1\left(m_2^2-m_1^2\right)}
y_{\alpha}(\tau)\Dc_{\alpha}\Hc^2  \, \delta\Ac_{\mu}(T).
\end{equation} 
Performing the $y$ and $\psi$ integrals one obtains
\begin{eqnarray}
&& \hskip -2 cm 
\frac{i}{2}\left\langle \, \epsilon^{\mu\nu}
(m_1+m_2) \int_0^{\infty}\frac{dT}{T}\int_0^Td\tau_1e^{-T m_1^2-\tau_1\left(m_2^2-m_1^2\right)}
\dot{g}_B(T,\tau_1) \,  \Dc_{\mu}\Hc  \,\delta\Ac_{\nu}
\right\rangle \nonumber\\
 &=&
\frac{i}{2}\left\langle  \epsilon^{\mu\nu}
\,J^2_{12}(m_1+m_2) \, \Dc_{\mu}\Hc \, \delta\Ac_{\nu}
\right\rangle.
\end{eqnarray} 
The second term of the insertion does not contribute at lowest order
since it is already of second order in derivatives but lacks the
appropriate fermionic factor to saturate the zero modes. The third
term of the insertion leads only to one contribution of the form
\begin{equation}
\label{C2dlo2c}
-2\psi_{\mu}\left\lbrace\delta\Ac_{\mu},\Hc\right\rbrace
\int_0^Td\tau_1e^{-T m_1^2-\tau_1\left(m_2^2-m_1^2\right)}
\left(-2i\,\psi_{\nu}\psi_5\Dc_{\nu}\Hc\right).
\end{equation} 
yielding
\begin{eqnarray}
&& \hskip -2 cm
-\frac{i}{2}\left\langle \,\epsilon^{\mu\nu}
(m_1-m_2)\int_0^{\infty}\frac{dT}{T}
\int_0^Td\tau_1e^{-T m_1^2-\tau_1\left(m_2^2-m_1^2\right)}\, \Dc_{\mu}\Hc  \,\delta\Ac_{\nu}
\right\rangle\nonumber\\
&=& -\frac{i}{2}\left\langle \epsilon^{\mu\nu} \,J^1_{12}(m_1-m_2)\Dc_{\mu}\Hc \, \delta\Ac_{\nu}
\right\rangle.
\end{eqnarray}
The factor $(m_1-m_2)$  results from the anticommutator in
\eq~(\ref{C2dlo2c}), and the sign change in the cyclic property of the 
trace as explained in \eq~(\ref{norm_ex}). The integrals $J$ are given
in Appendix \ref{app:Integrals}. The total current is hence given by
\be
\label{2dcovcont}
\delta W^{-}
= - i \left\langle \,\epsilon^{\mu\nu} 
A^1_{12}\Dc_{\mu}\Hc \,\delta\Ac_{\nu} \right\rangle, 
\ee
\be
A^1_{12} =  
 \frac{1}{m_1-m_2} - \frac{m_1m_2\log\left(m_1^2/m_2^2\right)}{(m_1-m_2)(m_1^2-m_2^2)},
\ee
where the function $A^1_{12}$ has been defined. 
This agrees with the results obtained in ref.~\cite{Salcedo2}.

\subsection{Effective Density}

The effective density can be obtained similar as the covariant
current, utilizing the insertion in
\eq~(\ref{insertionDensity}). Neglecting the antisymmetric tensor
$\Kc$, the insertion is
\begin{equation}
\label{insertionDensity2}
w(T)=-2i\,\psi_{\mu}\dot{x}_{\mu}\delta\Hc+2\psi_5\left[\delta\Hc,\Hc\right].
\end{equation} 

The contributions to the effective density are
\begin{eqnarray}
\label{cF2dlo}
\delta W^{-}&=&\left\langle\epsilon^{\mu\nu} \left(
\frac{i}{4}\left(J^1_{12}(m_1+m_2) + J^2_{12}(m_1-m_2)\right) \Fc_{\mu\nu} \right.\right.\nonumber\\
&&\left.\left. 
- \frac{1}{2}\left(J^5_{123}(m_1+m_3) + J^6_{123}(m_1+m_2) - J^7_{123}(m_2+m_3)\right) 
\Dc_{\mu}\Hc\Dc_{\nu}\Hc
\right) \delta\Hc \right\rangle\nonumber\\
&=& \left\langle \epsilon^{\mu\nu}
\left( -\frac{i}2 B^1_{12}\Fc_{\mu\nu}
 + B^2_{123}\Dc_{\mu}\Hc\Dc_{\nu}\Hc\right) \delta\Hc \right\rangle.
\end{eqnarray} 
where the functions $B_{12}$ and $B_{123}$ are given by
\begin{eqnarray}
B^1_{12}&=& - \frac{1}{m_1+m_2} -\frac{m_1m_2}
{(m_1+m_2)(m_1^2-m_2^2)}\log\left(\frac{m_1^2}{m_2^2}\right),\\
B^2_{123}&=&B_{123}^R+B_{123}^L\log(m_1^2)+
B_{\underline{23}1}^L\log(m_2^2)+B_{\underline{3}12}^L\log(m_2^2),
\end{eqnarray} 
with
\begin{equation}
B_{123}^R=\frac{1}{(m_1-m_2)(m_2-m_3)(m_1+m_3)},
\end{equation} 
\begin{equation}
B_{123}^L=\frac{(m_1^3+m_1m_2m_3)}{(m_1-m_2)(m_1+m_3)(m_1^2-m_2^2)(m_1^2+m_3^2)},
\end{equation} 
in accordance with ref.~\cite{Salcedo2}.

\subsection{Effective Action}

We proceed and briefly present the derivation of the imaginary part of
the effective action following ref.~\cite{Salcedo2}. Using the ansatz
in \eq~(\ref{ansatz_action}), the most general functional for $W_c^-$
consistent with chiral and gauge invariance in two dimensions reads
\begin{equation}
W_c^{-}=\left\langle \epsilon^{\mu\nu}N_{12}\Dc_{\mu}\Hc\Dc_{\nu}\Hc \right\rangle.
\end{equation} 
An additional term proportional to $\Fc$ could be added but it can be
removed by partial integration. Notice that $N_{12}$ is a real
function according to the comments made in the last section.

The function $N_{12}$ has some nontrivial restrictions. First of all,
the function $N_{12}$ is even in $m$ such that
\be
N(-m_1, -m_2) := N_{\underline{1}\underline{2}}= N_{12}.
\ee
Because of the cyclic property of the trace one obtains
\begin{equation}
N_{12}=N_{\underline{3}2}=N_{\underline{2}1}=N_{2\underline{1}},
\end{equation} 
and due to the Hermiticity of $W^{-}$
\begin{equation}
N_{12}=-N_{32}=-N_{\underline{1}2}=-N_{21}.
\end{equation} 
Varying $W_{c}^{-}[\Ac,\Hc]$ with respect to $\Ac$, one obtains
\begin{equation}
\delta W_c^{-}=
-i \left\langle\epsilon^{\mu\nu}  \left(-2\,(m_1+m_2)N_{12}\right) 
\Dc_{\mu} \Hc \delta\Ac_{\nu} \right\rangle.
\end{equation} 
Comparing this to \eq~(\ref{2dcovcont}) and adding the covariant
contribution in \eq~(\ref{Contr2d}) coming from $\Gamma_{gWZW}$ one
has
\begin{equation}
\frac{1}{m_1-m_2} - \frac{m_1m_2\log\left(m_1^2/m_2^2\right)}
{(m_1-m_2)(m_1^2-m_2^2)}=\frac{1}{2m_1}-\frac{1}{2m_2}-2\,(m_1+m_2)N_{12},
\end{equation} 
which finally leads to
\begin{equation}
\label{N12exp}
N_{12}=\frac12 \frac{m_1m_2}{m_1^2-m_2^2}
\left(\frac{\log(m_1^2/m_2^2)}{m_1^2-m_2^2}-\frac{1}{2}\left(\frac{1}{m_1^2}+
\frac{1}{m_2^2}\right)\right).
\end{equation} 
At higher order, the matching of the effective potential to the
current potentially becomes more intricate. On the other hand, the
anomaly only contributes to the leading order, such that the knowledge
of the covariant current (that in higher order coincides with the
consistent current) suffices to determine the effective action.

\subsection{Four Dimensions\label{4dim}}

For completeness, we also present the results for the effective action
and the effective current in four dimensions. The matching procedure
proceeds the same way as in ref.~\cite{Salcedo2}, and we do not repeat
it here.

The effective current in four dimensions consists out of three terms
and reads
\begin{equation}
\label{covcur4d}
\delta W_{d=4}^{-}= - i \left\langle\epsilon^{\mu\nu\lambda\sigma}
\left( - \frac{i}{2}A^2_{123}\Fc_{\nu\lambda}\Dc_{\mu}\Hc 
- \frac{i}{2}A^3_{123}\Dc_{\mu}\Hc\Fc_{\nu\lambda}
- \,A^4_{1234}\Dc_{\mu}\Hc\Dc_{\nu}\Hc\Dc_{\lambda}\Hc
\right)\delta\Ac_{\sigma} \right\rangle,
\end{equation}
while the effective density can be written as 
\begin{eqnarray}
\label{den4d}
\delta W_{d=4}^{-}\lefteqn{ = \left\langle \epsilon^{\mu\nu\lambda\sigma}\left(
\frac{1}{4}B^3_{123}\Fc_{\mu\nu}\Fc_{\lambda\sigma}
+\frac{i}{2}B^4_{1234}\Fc_{\lambda\sigma}\Dc_{\mu}\Hc\Dc_{\nu}\Hc\right.\right.}\nonumber\\
& & \quad  +\frac{i}{2}B^5_{1234}\Dc_{\mu}\Hc\Fc_{\lambda\sigma}\Dc_{\nu}\Hc
+\frac{i}{2}B^6_{1234}\Dc_{\mu}\Hc\Dc_{\nu}\Hc\Fc_{\lambda\sigma} \nn \\
&& \left. \left. \phantom{\frac12} - B^7_{12345}\Dc_{\mu}\Hc\Dc_{\nu}\Hc\Dc_{\lambda}
\Hc\Dc_{\sigma} \Hc \right) \delta\Hc \right\rangle.
\end{eqnarray} 
The functions $A^2_{123}$, $A^3_{123}$, $A^4_{1234}$, $B^3_{123}$,
$B^4_{1234}$, $B^5_{1234}$, $B^6_{1234}$, and $B^7_{12345}$ are given in
Appendix~\ref{app:ResEff}.

\section{Next to Leading Order Effective Action in Two Dimensions\label{sec_next}}

In this section we present as a novel result the imaginary part of the
effective action in next to leading order and two dimensions. Even
though the results are rather lengthy, the evaluation of the worldline
path integral involves only very basic integrals such that it can be
easily implemented using a computer algebra system.

In two dimensions and in next to leading order, the imaginary part of
the effective action takes the form
\begin{eqnarray}
\label{Wc}
W^c&=&\left\langle \epsilon^{\mu\nu}\left(
Q_{12} \Dc_{\mu} \Dc_{\alpha}\Hc\Dc_{\alpha} \Dc_{\nu}\Hc 
+\frac{i}{2}P_{12} \Fc_{\mu\nu}\Dc_{\alpha}\Dc_{\alpha}\Hc \right. \right. \nn \\
&& +\widetilde{R}_{123}\Dc_{\alpha}\Dc_{\alpha}\Hc\Dc_{\mu}\Hc\Dc_{\nu}\Hc
+\frac{i}{2}\widehat{R}_{123}\Fc_{\mu\nu}\Dc_{\alpha}\Hc\Dc_{\alpha}\Hc \nn \\
&& \left.\left.
+ M_{1234}\Dc_{\mu}\Hc\Dc_{\alpha}\Hc\Dc_{\nu}\Hc\Dc_{\alpha}\Hc\right)\right\rangle.
\end{eqnarray}
At next to leading order the action is chiral invariant and the
effective action can hence be immediately obtained by matching with
the covariant current that in this order coincides with the consistent
current. These functions must have the following properties
\be
\label{eq:PropP}
P_{12}=-P_{\underline{12}}=P_{21}, \quad 
Q_{12}=Q_{\underline{12}}=-Q_{21},
\ee
\be
\widetilde{R}_{123}=-\widetilde{R}_{\underline{123}}=\widetilde{R}_{\underline{21}3}, \quad
\widehat{R}_{123}=\widehat{R}_{\underline{123}}=-\widehat{R}_{\underline{21}3},
\ee
\be
M_{1234}=M_{\underline{1234}}=
-M_{\underline{34}12}=M_{4321}.
\ee

We have chosen a rather general imaginary effective action at the
required order which preserves gauge and chiral invariance, but we
have included a larger number of terms than necessary to perform the
matching process with the effective current. In fact, the matching
process could be done with solely the functions $Q_{12}$,
$\widetilde{R}_{123}$, $\widehat{R}_{123}$, and $M_{1234}$. Instead,
we have decided to include the additional term $P_{12}$, in order to
have the option of simplifying the action by a judicious choice of
this extra function. For example, the extra function can be used to
ensure that all functions remain finite at the coincidence limit,
as will be explained later on.

The calculation from the worldline formalism leads to the following
contributions to the covariant current
\begin{eqnarray}
\label{eq:WLcont}
\delta W^c&=&-i \, \epsilon^{\mu\nu} \left\langle 
I^1_{12}\Dc_{\mu}\Dc_{\alpha}\Dc_{\alpha}\Hc\delta\Ac_{\nu}
+ i\, I^2_{12}\Dc_{\alpha}\Fc_{\mu\alpha}\delta\Ac_{\nu}
+I^3_{123}\Dc_{\alpha}\Dc_{\alpha}\Hc\Dc_{\mu}\Hc\delta\Ac_{\nu}\right.\nonumber\\
&&\!\!+I^4_{123}\Dc_{\mu}\Hc\Dc_{\alpha}\Dc_{\alpha}\Hc\delta\Ac_{\nu}
+I^5_{123}\Dc_{\mu}\Dc_{\alpha}\Hc\Dc_{\alpha}\Hc\delta\Ac_{\nu}
+I^6_{123}\Dc_{\alpha}\Hc\Dc_{\mu}\Dc_{\alpha}\Hc\delta\Ac_{\nu}\nonumber\\
&&\!\!+ i\,  I^7_{123}\Fc_{\mu\alpha}\Dc_{\alpha}\Hc\delta\Ac_{\nu}
+ i \, I^8_{123}\Dc_{\alpha}\Hc\Fc_{\mu\alpha}\delta\Ac_{\nu}
+I^9_{1234}\Dc_{\mu}\Hc\Dc_{\alpha}\Hc\Dc_{\alpha}\Hc\delta\Ac_{\nu}\nonumber\\
&&\!\!\left.
+I^{10}_{1234}\Dc_{\alpha}\Hc\Dc_{\mu}\Hc\Dc_{\alpha}\Hc\delta\Ac_{\nu}
+I^{11}_{1234}\Dc_{\alpha}\Hc\Dc_{\alpha}\Hc\Dc_{\mu}\Hc\delta\Ac_{\nu}\right\rangle.
\end{eqnarray} 
The coefficient functions are given in Appendix \ref{curr_nlo}.  In
order to express the current in this form, partial integration has
been used to remove terms of the form $\Dc\delta\Ac$.  In addition,
indices that are contracted with the $\epsilon$ tensor have been moved
to the left, such that a term of the form $\Dc_{\alpha}\Dc_{\mu}$
yields a sum of terms of the type $\Dc_{\mu}\Dc_{\alpha}$ and
$\Fc_{\mu\alpha}$.

The contributions from the variation of \eq~(\ref{Wc}) can be grouped
in three levels, with the first level having only contributions from
$Q$ and $P$; the second level from the previous ones and
$\widetilde{R}$ and $\widehat{R}$; the last level with all
functions. Adding the contributions from the worldline method and the
variation of \eq~(\ref{Wc}) one obtains for the first level the
following two equations
\begin{eqnarray}
\label{eq:1stLevel}
\,P_{\underline{2}1} + (m_1+m_2)Q_{12} &=&I^1_{12},\nonumber\\
(m_1 + m_2)P_{12}-(m_1^2-m_2^2)Q_{\underline{2}1} &=&I^2_{12},
\end{eqnarray} 
which have the solution
\begin{eqnarray}
\label{Q}
Q_{12}&=&\frac{I^{2}_{\underline{2}1}}{m_1^2-m_2^2}-
\frac{P_{\underline{2}1}}{m_1+m_2}.
\end{eqnarray} 
Besides, there arises the following restriction which is satisfied and
can serve as a check for the corresponding terms in the effective
current
\be
(m_1+m_2)I^1_{\underline{2}1} = - I^2_{12}, \quad 
I^1_{12}= - I^1_{\underline{12}}, \quad 
I^2_{12}=I^2_{\underline{12}}.
\ee
The matching equations for the next level are
\begin{eqnarray}
\label{eq:2ndLevela}
- \nabla_{2}\left((m_1 + m_2) Q_{\underline{2}1} -P_{\underline{2}1}\right)
+ Q_{12} + Q_{\underline{2}1} 
+(m_1+m_3)( -\widetilde{R}_{123} +\widetilde{R}_{\underline{3}12})&=&I^3_{123},\\
\label{eq:2ndLevelb}
\nabla_{2}\left((m_1 + m_2)(Q_{12}+Q_{\underline{2}1}) \right)
-(m_1+m_3)\widetilde{R}_{\underline{3}12}
-2 Q_{12} -2 Q_{\underline{2}1} 
+ \widehat{R}_{\underline{3}12}&=&I^5_{123},\\
\label{eq:2ndLevelc}
\nabla_{2}\left((m_1 + m_2)P_{12} \right) 
- (m_1 - m_2)\nabla_{2}\left((m_1+ m_2) Q_{\underline{2}1} \right)  && \nn \\  
-2 P_{12} + 2 (m_1 - m_2) Q_{\underline{2}1} 
+ (m_1 + m_3)(Q_{13} + 2 Q_{\underline{3}1})&&\nonumber\\
- \widehat{R}_{123} (m_1 + m_3)  
+(m_1 - m_2) (m_1 + m_3) \widetilde{R}_{\underline{3}12}  &=&I^7_{123},
\end{eqnarray} 
and their complex conjugates.

The first \eq~(\ref{eq:2ndLevela}) is of the form 
\be
\widetilde{R}_{123} - \widetilde{R}_{\underline{3}12}
=- \frac{ \tilde I^3_{123}}{m_1+m_3},
\ee
and a set of solutions to \eqs~(\ref{eq:2ndLevela}) and
(\ref{eq:2ndLevelb}) is hence given by
\begin{eqnarray}
\label{eq:SolRRR}
\widetilde{R}_{123}&=&-\frac{1}{2}
\left(\frac{\widetilde{I}_{123}}{m_1+m_3}\right)_{123}
-\frac{1}{2}\left(\frac{\widetilde{I}_{123}}{m_1+m_3}\right)_{\underline{3}12}
-\frac{1}{2}\left(\frac{\widetilde{I}_{123}}{m_1+m_3}\right)_{\underline{23}1}\\
\label{eq:SolRRRR}
\widehat{R}_{123}&=&\widehat{I}_{\underline{23}1}-(m_1-m_2)\widetilde{R}_{123}.
\end{eqnarray}
The functions $\widetilde{I}$ and $\widehat{I}$ are hereby defined as
\begin{eqnarray}
\label{eq:widetildeI}
\widetilde{I}_{123}&=&I^3_{123}
+ \nabla_{2}\left((m_1 + m_2) Q_{\underline{2}1} - P_{\underline{2}1}\right)
- Q_{12}- Q_{\underline{2}1}, \\
\label{eq:widehatI}
\widehat{I}_{123}&=&I^5_{123}
-\nabla_{2}\left((m_1 + m_2)(Q_{12}+Q_{\underline{2}1})\right)
+ 2 Q_{12}+ 2 Q_{\underline{2}1}.
\end{eqnarray} 
The last \eq~(\ref{eq:2ndLevelc}) leads to a constraint on the $I$
functions that is given in Appendix~\ref{curr_nlo}.

The function $\widetilde{R}$ possesses the required symmetries, and it
reproduces the effective current correctly, but it is not necessarily
finite in the coincidence limit, $m_1 \to -m_3$. One way of solving
this problem is to choose the function $P$ appropriately which up to
this point remained undetermined. Such a choice is e.g. given by
\be
P_{12} = \frac{I^2_{12}}{m_1+m_2}, \quad Q_{12} = 0,
\ee
which leaves $\widetilde{I}$ as
\bea
\label{eq:widetildeIb}
\widetilde{I}_{123} &=&I^3_{123}
-\frac{I^2_{\underline{2}1}}{(m_1-m_2)(m_2-m_3)}
+\frac{I^2_{\underline{3}1}}{(m_1-m_3)(m_2-m_3)}.
\eea 
With this choice, $\widetilde{R}$ is finite in the coincidence limit,
as can be checked explicitly, and since $\widehat{I}$ is also finite,
so is $\widehat{R}$.

For the last level, the following three equations hold
\begin{eqnarray}
\label{eqM1}
\nabla_{1} \widehat{R}_{\underline{3}12} 
- (\nabla_{2}+\nabla_{3})
\left((m_1 + m_3) \widetilde{R}_{\underline{3}12} \right) 
 + 2 \widetilde{R}_{\underline{3}12}  &&\nonumber\\ 
- 2 (m_1 + m_4) M_{1234}  &=&I^{9}_{1234},\\
\label{eqM2}
 ( \nabla_{3} - \nabla_1 )\left(  \widetilde{R}_{\underline{3}12} (m_1 + m_3)\right)
+ \nabla_{2} \widehat{R}_{\underline{3}12} 
- 2  \widetilde{R}_{\underline{3}12} + 2 \widetilde{R}_{\underline{4}23}
&&\nonumber\\
-2 (m_1 + m_4) M_{\underline{4}123}  &=&I^{10}_{1234},\\ 
\label{eqM3}
(\nabla_{1}+\nabla_{2}) \left((m_1 + m_3) \widetilde{R}_{\underline{3}12} \right)
 + \nabla_{3} \widehat{R}_{\underline{3}12} 
- 2 \widetilde{R}_{\underline{4}23} &&\nonumber\\
+2 (m_1 + m_4) M_{1234})&=&I^{11}_{1234}.
\end{eqnarray} 

One of these equations can be used to determine $M$, while the other
two lead again to constraints on the $I$ functions. the sum of the
three equations has the especially simple form
\begin{eqnarray} 
-2 (m_1 + m_4 ) M_{\underline{4}123}&=& 
- (\nabla_{1} + \nabla_{2} + \nabla_{3}  ) \widehat{R}_{\underline{3}12} 
+ I^{9}_{1234} + I^{10}_{1234} + I^{11}_{1234}. 
\end{eqnarray} 

Since all previous functions in the effective action have been chosen
finite in the coincidence limit, so is $M_{1234}$.
\eqs~(\ref{eqM1}) and (\ref{eqM3}) show that $M_{1234}$ is finite in
the limit $m_1\to m_2$, while \eq~(\ref{eqM2}) shows that $M_{1234}$
is finite in the limit $m_1\to -m_4$. This concludes the discussion of
the next to leading order contributions in two dimensions.

\section{Conclusions}

We presented a worldline formalism to calculate the imaginary part of
the covariant current of a general chiral model in the derivative
expansion and our results are best summarized by \eqs~(\ref{dWIm2})
and (\ref{ohneK}).

The resulting covariant current can be used to reproduce the imaginary
part of the effective action by integration or matching. The advantage
of this approach, compared to explicit formulas of the effective
action, as given e.g. in ref.~\cite{Gagne2}, consists in the manifest
chiral covariance. Chiral covariance reduces the number of possible
contributions to the current enormously and makes even next to leading
order calculations manageable as demonstrated in section
\ref{sec_next} in the case of two dimensions. Besides the chiral
covariance, the use of the worldline formalism is essential in our
approach. The evaluation of the worldline path integral requires
neither performing Dirac algebra nor integrating over momentum space,
in contrast to the more traditional methods used in
ref.~\cite{Salcedo2}.

In principle, it is possible to use the presented formalism to
determine the effective CP violation resulting from integrating out
the fermions of the Standard model. For example, in next to leading
order in four dimensions, an operator of the form $\Dc\Hc \Dc\Hc \Fc
\Fc$ could arise from the CP violation in the CKM matrix. Since the
mass terms of the fermions are treated non-perturbatively in the
derivative expansion, the resulting effective CP-violating operator is
not necessarily proportional to the Jarlskog determinant.  The
discussion of this question is postponed to a forthcoming publication.

\section*{Acknowledgments}

T.K. is supported by the Swedish Research Council
(Vetenskapsr{\aa}det), Contract No.~621-2001-1611. A.H. is supported
by CONACYT/DAAD, Contract No.~A/05/12566.

\appendix

\section{Integrals used in the calculation \label{app:Integrals}}

In this section, the function $g$ denotes the bosonic Green function
\be
g(T, \tau_1) = \langle y(T) y(\tau_1)\rangle,
\ee
and
\be
\dot g( T, \tau_1 ) = \langle \dot y(T) y(\tau_1)\rangle 
= - 2 \langle \psi_A(T) \psi_A(\tau_1)\rangle,
\ee
where the last expression does not contain a summation over the index
$A$.

\subsection{Integrals in Two Dimensions \label{app:Integrals2dim}}

In the calculation in two dimensions the following integrals have
been used
\bea
J^1_{12}&=&\int^{\infty}_0 \frac{dT}{T}\int^T_0 d\tau_1 \,e^{-T m^2_1-\tau_1(m^2_2-m^2_1)}
=\frac{\log(m_1^2/m_2^2)}{m_1^2-m_2^2},\nn\\
J^2_{12}&=&\int^{\infty}_0 \frac{dT}{T}\int^T_0 dt_1 \,e^{-T m^2_1-\tau_1(m^2_2-m^2_1)}
\dot g(T, \tau_1)\nn\\
&=&-\frac{2}{m_1^2-m_2^2}+\frac{(m_1^2+m_2^2)}{(m_1^2-m_2^2)^2}\log\left(\frac{m_1^2}{m_2^2}\right),\nn\\
J^{3}_{12}&=&\int^{\infty}_0 \frac{dT}{T} \int^T_0 dt_1e^{-T m^2_1-\tau_1(m^2_2-m^2_1)}
\dot g (T, \tau_1) g (T, \tau_1)\nonumber\\
&=&-3\frac{m_1^2+m_2^2}{(m_1^2-m_2^2)^3}
+\frac{(m_1^4+4m_1^2m_2^2+m_2^4)}{(m_1^2-m_2^2)^4}\log\left(\frac{m_1^2}{m_2^2}\right). 
\eea
The remaining occurring integrals can be expressed as
\bea
J^4_{12}&=&\int^{\infty}_0 \frac{dT}{T}\int^T_0 d\tau_1  \,e^{-T m^2_1-\tau_1(m^2_2-m^2_1)}
g(T, \tau_1)=\frac{J^2_{12}}{m_1^2-m_2^2},\nn\\
J^5_{123}&=&\int^{\infty}_0 \frac{dT}{T}\int^T_0 d\tau_1 
\int_0^{\tau_1}d\tau_2  \,e^{-T m^2_1- \sum_{i=1}^2 \tau_i(m^2_{i+1}-m^2_i) }=
-\frac{\nabla_2 J^1_{12}}{m_2+m_3},\nn\\
J^6_{123}&=&\int^{\infty}_0 \frac{dT}{T}\int^T_0 d\tau_1
\int_0^{\tau_1}d\tau_2  \,e^{-T m^2_1- \sum_{i=1}^2 \tau_i(m^2_{i+1}-m^2_i) } 
\,\dot g(T, \tau_1) = -\frac{\nabla_2 J^2_{12}}{m_2+m_3},\nn\\
J^7_{123}&=&\int^{\infty}_0 \frac{dT}{T}\int^T_0 d\tau_1
\int_0^{\tau_1}d\tau_2  \,e^{-T m^2_1- \sum_{i=1}^2 \tau_i(m^2_{i+1}-m^2_i) }  
\, \dot g(T, \tau_2)=-J^6_{321}.
\eea

\subsection{Integrals in Four Dimensions \label{app:Integrals4dim}}
In four dimensions the following integrals with three indices are used
\bea
J^8_{123}&=&\int_0^{\infty}\frac{dT}{T^2}\int_0^Td\tau_1\int_0^{\tau_1}d\tau_2
\,e^{-T m_1^2- \sum_{i=1}^2 \tau_i(m^2_{i+1}-m^2_i) }\nonumber\\
&=&\frac{m_1^2\log(m_1^2)}{(m_1^2-m_2^2)(m_1^2-m_3^2)}-
\frac{m_2^2(m_1^2-m_3^2)\log(m_2^2)}{(m_1^2-m_2^2)(m_1^2-m_3^2)(m_2^2-m_3^2)}\nonumber\\
&&+\frac{m_3^2\log(m_3^2)}{(m_1^2-m_3^2)(m_2^2-m_3^2)}, \nn \\
J^9_{123}&=&\int_0^{\infty}\frac{dT}{T^2}\int_0^Td\tau_1\int_0^{\tau_1}d\tau_2
\,e^{-T m_1^2- \sum_{i=1}^2 \tau_i(m^2_{i+1}-m^2_i) } \, \dot g (T,\tau_1)\nonumber\\
&=&-\frac{m_1^2}{(m_1^2-m_2^2)(m_1^2-m_3^2)}
+\frac{m_1^2(m_1^4-m_2^2m_3^2)\log(m_1^2)}{(m_1^2-m_2^2)^2(m_1^2-m_3^2)^2}\nonumber\\
&&-\frac{m_1^2m_2^2\log(m_2^2)}{(m_1^2-m_2^2)^2(m_2^2-m_3^2)}
+\frac{m_1^2m_3^2\log(m_3^2)}{(m_1^2-m_3^2)^2(m_2^2-m_3^2)}, \nn \\
J^{10}_{123}&=&\int_0^{\infty}\frac{dT}{T^2}\int_0^Td\tau_1\int_0^{\tau_1}d\tau_2
\, e^{-T m_1^2- \sum_{i=1}^2 \tau_i(m^2_{i+1}-m^2_i) }
\, \dot g(T,\tau_2)=-J^{9}_{321}.
\eea
The integrals with four indices can be expressed as
\bea
J^{11}_{1234}&=&\int_0^{\infty}\frac{dT}{T^2}
\int_0^Td\tau_1\int_0^{\tau_1}d\tau_2\int_0^{\tau_2}d\tau_3
\, e^{-T m_1^2- \sum_{i=1}^3 \tau_i(m^2_{i+1}-m^2_i)  }\nonumber\\
&=&-\frac{\nabla_3J^8_{123}}{m_3+m_4}, \nn\\
J^{12}_{1234}&=&\int_0^{\infty}\frac{dT}{T^2}
\int_0^Td\tau_1\int_0^{\tau_1}d\tau_2\int_0^{\tau_2}d\tau_3
e^{-T m_1^2-  \sum_{i=1}^3 \tau_i(m^2_{i+1}-m^2_i) }
\, \dot g (T,\tau_1) \nonumber\\ &=&-\frac{\nabla_3 J^9_{123}}{m_3+m_4}, \nn \\
J^{13}_{1234}&=&\int_0^{\infty}\frac{dT}{T^2}
\int_0^Td\tau_1\int_0^{\tau_1}d\tau_2\int_0^{\tau_2}d\tau_3
 e^{-T m_1^2- \sum_{i=1}^3 \tau_i(m^2_{i+1}-m^2_i)  } 
\, \dot g (T,\tau_2)\nonumber\\
&=&-\frac{\nabla_3 J^{10}_{123}}{m_3+m_4}, \nn \\
J^{14}_{1234}&=&\int_0^{\infty}\frac{dT}{T^2}
\int_0^Td\tau_1\int_0^{\tau_1}d\tau_2\int_0^{\tau_2}d\tau_3
e^{-T m_1^2- \sum_{i=1}^3 \tau_i(m^2_{i+1}-m^2_i)  }
\, \dot g (T,\tau_3) \nn \\
&=&-J^{12}_{4123}.
\end{eqnarray} 
Finally, the integrals with five indices read
\begin{eqnarray}
J^{15}_{12345} &=&\int_0^{\infty}\frac{dT}{T^2}
\int_0^Td\tau_1\int_0^{\tau_1}d\tau_2\int_0^{\tau_2}d\tau_3\int_0^{\tau_3}d\tau_4 
\, e^{-T m_1^2- \sum_{i=1}^4 \tau_i(m^2_{i+1}-m^2_i) } \nonumber\\
&=&-\frac{\nabla_4 J^{11}_{1234} }{m_4+m_5}, \nn \\
J^{16}_{12345} &=&\int_0^{\infty}\frac{dT}{T^2}
\int_0^Td\tau_1\int_0^{\tau_1}d\tau_2\int_0^{\tau_2}d\tau_3\int_0^{\tau_3}d\tau_4 
\, e^{-T m_1^2- \sum_{i=1}^4 \tau_i(m^2_{i+1}-m^2_i) } \dot{g}(T,\tau_1)\nonumber\\
&=&-\frac{\nabla_4 J^{12}_{1234} }{m_4+m_5}, \nn \\
J^{17}_{12345} &=&\int_0^{\infty}\frac{dT}{T^2}
\int_0^Td\tau_1\int_0^{\tau_1}d\tau_2\int_0^{\tau_2}d\tau_3\int_0^{\tau_3}d\tau_4 
\, e^{-T m_1^2- \sum_{i=1}^4 \tau_i(m^2_{i+1}-m^2_i) } \dot{g}(T,\tau_2)\nonumber\\
&=&-\frac{\nabla_4 J^{13}_{1234} }{m_4+m_5}, \nn \\
J^{18}_{12345} &=&\int_0^{\infty}\frac{dT}{T^2}
\int_0^Td\tau_1\int_0^{\tau_1}d\tau_2\int_0^{\tau_2}d\tau_3\int_0^{\tau_3}d\tau_4 
\, e^{-T m_1^2- \sum_{i=1}^4 \tau_i(m^2_{i+1}-m^2_i) } \dot{g}(T,\tau_3)\nonumber\\
&=&-\frac{\nabla_4 J^{14}_{1234} }{m_4+m_5}, \nn \\
J^{19}_{12345} &=&\int_0^{\infty}\frac{dT}{T^2}
\int_0^Td\tau_1\int_0^{\tau_1}d\tau_2\int_0^{\tau_2}d\tau_3\int_0^{\tau_3}d\tau_4 
\,  e^{-T m_1^2- \sum_{i=1}^4 \tau_i(m^2_{i+1}-m^2_i) } \dot{g}(T,\tau_4)\nonumber\\
&=&-\frac{\nabla_3 J^{12}_{1234} }{m_3+m_4}.
\end{eqnarray}

\section{Results in four dimensions \label{app:ResEff} } 

In this section, we give the coefficient functions for the
effective current and the effective density in four dimensions
introduced in section~\ref{4dim}.

The functions of the covariant current are given by
\begin{eqnarray}
A^2_{123}&=&\frac{m_1m_2-m_1m_3-m_2m_3}{(m_1+m_2)(m_1-m_3)(m_2-m_3)} \nn \\
&& +\frac{m_1^3(m_1(m_2-m_3)-2m_2m_3)\log[m_1^2/m_3^2]}
{(m_1+m_2)(m_1-m_3)(m_1^2-m_2^2)(m_1^2-m_3^2)}\nonumber\\
&&
+\frac{m_2^3(m_2(m_3-m_1)+2m_1m_3)\log[m_2^2/m_3^2]}
{(m_1+m_2)(m_2-m_3)(m_1^2-m_2^2)(m_2^2-m_3^2)}, \\
A^3_{123}&=&-A^1_{321},
\eea
and
\begin{equation}
A^4_{1234}=A_{1234}^{R}+A_{1234}^{L}\log[m_1^2]+A_{2341}^{L}\log[m_2^2]+A_{3412}^{L}\log[m_3^2]
+A_{4123}^{L}\log[m_4^2],
\end{equation} 
with
\begin{eqnarray}
A_{1234}^{R}&=&\frac{m_1m_2m_3-m_1m_2m_4+m_1m_3m_4-m_2m_3m_4}
{(m_1-m_2)(m_1+m_3)(m_1-m_4)(m_2-m_3)(m_2 + m_4)(m_3-m_4)},\nonumber\\
A_{1234}^{L}&=&\frac{m_1^3(-m_1^3-m_1m_2m_3+m_1m_2m_4-m_1m_3m_4+2m_2m_3m_4)}
{(m_1-m_2)(m_1+m_3)(m_1-m_4)(m_1^2-m_2^2)(m_1^2-m_3^2)(m_1^2-m_4^2)}.
\end{eqnarray}

The explicit functions occurring in the effective density are rather
lengthy and hence we display them in terms of the integrals presented
in the last section.
\bea
B^3_{123} &=& J^8_{123} (m_1 + m_5) 
- J^9_{123} (m_1 - m_3) - J^{10}_{123} (m_2 - m_3), \nn \\
B^4_{1234} &=& J^{11}_{1234} (m_1 + m_4) - J^{12}_{1234} (m_1 - m_2) 
- J^{13}_{1234} (m_2 + m_3) + J^{14}_{1234} (m_4 + m_3), \nn \\
B^5_{1234} &=& J^{11}_{1234} (m_1 + m_4) - J^{12}_{1234} (m_1 + m_2) 
- J^{13}_{1234} (m_2 - m_3) + J^{14}_{1234} (m_4 + m_3), \nn \\
B^6_{1234} &=& B^4_{4321}, \nn \\
B^7_{12345} &=& J^{15}_{12345} (m_1 + m_5) -  J^{16}_{12345} (m_1 + m_2)
+ J^{17}_{12345} (m_2 + m_3) \nn \\
&& -  J^{18}_{12345} (m_3 + m_4) + J^{19}_{12345} (m_4 + m_5).
\eea

\section{Covariant Current in Next to Leading Order\label{curr_nlo}}
In this section we summarize the contributions to the covariant
current in next to leading order. For the first two levels, they are
given by
\bea
I^1_{12}&=&\frac{I^2_{\underline{2}1}}{m_1-m_2}, \quad
I^2_{12}=-4(m^2_1-m^2_2)J^3_{12}+4(m_1+m_2)^2 J^4_{12}, \nn \\
I^3_{123}&=& \frac{m_1-m_2+2m_3}{m^2_2-m^2_3}\nabla_2(I^2_{\underline{2}1})
+8\frac{m_1+m_3}{m^2_2-m^2_3}\nabla_2(m_1m_2J^4_{12})\nonumber\\
&& +16\frac{m_1m_3J^4_{13}}{m^2_2-m^2_3}
-\frac{I^2_{\underline{3}1}}{(m_1-m_3)(m_2-m_3)}, \nn \\
I^5_{123}&=& 2\frac{\nabla_2(I^2_{\underline{2}1})}{m_1-m_2}
+3 \frac{I^2_{\underline{3}1}}{(m_1-m_3)(m_1-m_2)}-I^3_{123}
-\frac{I^7_{\underline{1}32}}{m_1-m_2}, \nn \\
I^7_{123}&=&(m_2+m_3)\left(\left(\frac{\nabla_1I^2_{12}}{m_1+m_2}\right)_{\underline{3}12}-\frac{\nabla_1I^2_{12}}{m_1+m_2}\right)
+\frac{3m_2-m_3}{m_2+m_3}\nabla_2I^2_{12}\nonumber\\
&&+4(m_2+m_3)(m_1(m_1-m_2-m_3)+m_2m_3)\left(\left(\frac{\nabla_1J^4_{12}}{m_1+m_2}\right)_{\underline{3}12}-\frac{\nabla_1J^4_{12}}{m_1+m_2}\right)\nonumber\\
&&-8(m_2+m_3)\left(\left(\frac{\nabla_1(m_1m_2J^4_{12})}{m_1+m_2}\right)_{\underline{3}12}-\frac{\nabla_1(m_1m_2J^4_{12})}{m_1+m_2}\right)\nonumber\\
&&+4(m_2-m_3)\left(\frac{\nabla_2\left((m_1^2-4m_1m_2+m_2^2)J^4_{12}\right)}{m_2+m_3}-\nabla_2\left((m_1-m_2)J^4_{12}\right)\right),
\eea
and the relations
\be
I^4_{123}=-I^3_{321}, \quad
I^6_{123}=-I^5_{321}, \quad
I^8_{123}=I^7_{321}. 
\ee
The contributions to the last level read
\begin{eqnarray}
I^{9}_{1234}&=&-I^{11}_{4321}, \nn \\
I^{10}_{1234}&=&-\frac{m_1+m_4}{m_3-m_4}I^{11}_{\underline{4}123}
+\frac{1}{2}\left(\nabla_2+\frac{m_1+m_4}{m_3-m_4}
\left(\nabla_3\right)_{\underline{4}123} \right)I^5_{123}\nonumber\\
&&+\left(\nabla_2+\nabla_3+\frac{m_1+m_4}{m_3-m_4}
\left(\nabla_3\right)_{\underline{4}123}\right)\frac{m_1+m_3}{m_2-m_3}
(I^3_{\underline{21}3}-(\nabla_2 I^1_{12})_{\underline{21}3})\nonumber\\
&&+\frac{1}{2}\left(\nabla_2+\nabla_3+\frac{m_1+m_4}{m_3-m_4}
\left(-\nabla_1+\nabla_2+\nabla_3\right)_{\underline{4}123} \right)
(I^3_{123}-\nabla_2 I^1_{12})\nonumber\\
&&+\frac{1}{2}\left(-\nabla_1+\nabla_2+\frac{m_1+m_4}{m_3-m_4}
\left(\nabla_3\right)_{\underline{4}123}\right)
(I^3_{321}-(\nabla_2 I^1_{12})_{321})\nonumber\\
&&-\frac{m_1+m_3}{2(m_2-m_3)(m_3-m_4)}(I^3_{\underline{21}3}-(\nabla_2 I^1_{12})_{\underline{21}3})
+\frac{2}{m_2+m_4}(I^3_{432}-(\nabla_2I^1_{12})_{432})\nonumber\\
&&-\frac{1}{2(m_3-m_4)}(I^3_{421}-(\nabla_2I^1_{12})_{421}), \nn \\
I^{11}_{1234}&=&\nabla_2 I^4-\frac{I^4_{124}}{m_3-m_4}
-\frac{4(m_1^2-m_3^2)J^3_{13}}{(m_3-m_4)(m_1^2-m_4^2)}
-\frac{4(m_3+m_4)J^4_{34}}{m_1^2-m_4^2}-\frac{4J^5_{124}}{m_3-m_4}\nonumber\\
&&-\frac{m_1+m_2}{m_3-m_4}\left(\nabla_2-\frac{m_2+m_3}{m_3+m_4}\nabla_3\right)
\left(\frac{\nabla_1((m_1^2-m_2^2)(I^2_{12}-8m_1 m_2 J^4_{12}))}
{(m_1^2-m_2^2)(m_1+m_2)}\right)\nonumber\\
&&+\frac{(m_3+m_4)\left(\nabla_1(I^2_{12}-8m_1 m_2 
J^4_{12})\right)_{\underline{4}13}}{(m_1^2-m_4^2)(m_1-m_4)}
+\frac{4(m_2+m_3)(m_1-m_4)}{m_1+m_2)(m_3-m_4)}\nabla_1J^4\nonumber\\
&&-\frac{4(m_1+m_2)}{m_3-m_4}\left(-\nabla_2+\frac{m_2+m_3}{m_3+m_4}\nabla_3\right)
\frac{(m_1^2+m_3^2)J^4_{13}-(m_2^2-m_3^2)J^4_{23}}{m_1^2-m_2^2}\nonumber\\
&&-\frac{4(m_2+m_3)(m_1-m_4)(m_3+m_4)}{m_1+m_2}
\nabla_1\frac{\left(\nabla_1J^4\right)_{\underline{3}12}}{m_1-m_3}
-\frac{4(m_1+m_3)}{m_3+m_4}\nabla_3J^5\nonumber\\
&&-\frac{4(m_1+m_2)(m_2+m_4)}{m_3^2-m_4^2}\nabla_3 J^6. 
\end{eqnarray}

All functions $I$ are finite in the coincidence limit and must fulfill
the following constraints due to the behavior of the terms in the
effective action under cyclic permutation and complex conjugation:
\begin{eqnarray}
\label{current_constraints}
\frac{(m_1+m_3)I^3_{\underline{3}12}}{m_2-m_3}+I^3_{321}
+\frac{I^2_{13}}{(m_1-m_2)(m_2-m_3)}
+\frac{I^2_{\underline{3}1}}{(m_1-m_2)(m_1-m_3)}&&\nonumber\\
-\frac{(m_1+m_3)I^2_{23}}{(m_1-m_2)(m_2^2-m_3^2)}-
\frac{I^2_{\underline{3}2}}{(m_1-m_2)(m_2-m_3)}&=&0,\nonumber\\
I^3_{123}+I^3_{321}-\frac{(m_1+m_3)I^3_{\underline{23}1}}{m_1-m_2}+I^5_{123}+I^5_{321}&&\nonumber\\
+\frac{I^2_{12}}{m_1^2-m_2^2}+
\frac{I^2_{\underline{2}1}}{(m_1-m_2)(m_2-m_3)}
-\frac{I^2_{\underline{3}1}}{(m_1-m_2)(m_2-m_3)}&=&0,\nonumber\\
(m_1+m_3) I^3_{\underline{3}12}+(m_1+m_3) I^5_{\underline{3}12}-I^8_{123}
+\frac{2I^2_{13}}{m_1-m_2}&&\nonumber\\
+\frac{(3 m_1-m_2+2 m_3) I^2_{23}}{(m_1-m_2)(m_2+m_3)}&=&0, 
\eea
and 
\bea
I^3_{123} = I^3_{\underline{123}}=0, \quad
I^5_{123} = I^5_{\underline{123}}=0, \quad
I^7_{123} = -I^7_{\underline{123}}=0.
\end{eqnarray} 

\bibliographystyle{unsrt}

\bibliography{bibliography}

\end{document}